\documentclass[fleqn,usenatbib]{mnras}

\usepackage{newtxtext,newtxmath}

\usepackage[T1]{fontenc}

\DeclareRobustCommand{\VAN}[3]{#2}
\let\VANthebibliography\thebibliography
\def\thebibliography{\DeclareRobustCommand{\VAN}[3]{##3}\VANthebibliography}

\usepackage{graphicx}	
\usepackage{amsmath}	
\usepackage{booktabs}
\usepackage{color}
\usepackage{caption}
\usepackage{subcaption}
\usepackage{multicol,multirow}

\title[Classifying supernovae from multiple surveys]{Pan-chromatic photometric classification of supernovae from multiple surveys and transfer learning for future surveys}

\author[Burhanudin \& Maund]{
Umar. F. Burhanudin,$^{1}$\thanks{E-mail: ufburhanudin1@sheffield.ac.uk}
Justyn. R. Maund$^{1}$
\\
$^{1}$Department of Physics and Astronomy, University of Sheffield, Sheffield S3 7RH, UK\\
}

\date{Accepted XXX. Received YYY; in original form ZZZ}

\pubyear{2022}

\begin{document}
\label{firstpage}
\pagerange{\pageref{firstpage}--\pageref{lastpage}}
\maketitle

\begin{abstract}
Time-domain astronomy is entering a new era as wide-field surveys with higher cadences allow for more discoveries than ever before. The field has seen an increased use of machine learning and deep learning for automated classification of transients into established taxonomies. Training such classifiers requires a large enough and representative training set, which is not guaranteed for new future surveys such as the Vera Rubin Observatory, especially at the beginning of operations. We present the use of Gaussian processes to create a uniform representation of supernova light curves from multiple surveys, obtained through the Open Supernova Catalog for supervised classification with convolutional neural networks. We also investigate the use of transfer learning to classify light curves from the Photometric LSST Astronomical Time Series Classification Challenge (PLAsTiCC) dataset. Using convolutional neural networks to classify the Gaussian process generated representation of supernova light curves from multiple surveys, we achieve an AUC score of 0.859 for classification into Type Ia, Ibc, and II. We find that transfer learning improves the classification accuracy for the most under-represented classes by up to 18\% when classifying PLAsTiCC light curves, and is able to achieve an AUC score of 0.945 when including photometric redshifts for classification into six classes (Ia, Iax, Ia-91bg, Ibc, II, SLSN-I). We also investigate the usefulness of transfer learning when there is a limited labelled training set to see how this approach can be used for training classifiers in future surveys at the beginning of operations.
\end{abstract}




\section{Introduction}

The emergence of synoptic all-sky surveys with increased coverage of the night sky (both in area and time) has allowed astronomers to discover more objects more quickly (e.g. Pan-STARRS; \citealt{kaiser2010PS1}, Asteroid Terrestrial-impact Last Alert System; \citealt{tonry2018}, All Sky Automated Survey for SuperNovae; \citealt{shappee2014}, the Gravitational-wave Optical Transient Observer; \citealt{steeghs2021}, Zwicky Transient Facility; \citealt{bellm2019ztf}). The rate of discovery and data collection of current surveys and that of expected future surveys such as  the Legacy Survey of Space and Time (LSST; \citealt{lsst2019}) on the Vera Rubin Observatory has prompted work on machine learning and deep learning approaches to automate the identification and classification of new transients. The motivation for photometric classification of supernovae arises from the fact that not all discovered supernovae will be subject to spectroscopic follow-up for spectral classification. The ability to classify supernovae based on just light curves will benefit the studies of cosmology with Type Ia supernovae (\citealt{1998riess}, \citealt{1999Perlmutter}, \citealt{betoule2014}) and accumulating a large sample of core-collapse supernovae allows for population studies to understand their diversity (e.g. \citealt{modjaz2019}).  In the past decade, a lot of work has been done on supernova light curve classification using photometric observations with machine learning and deep learning. At present, most of these studies focus on classifying supernovae from a single survey with either real or simulated data.

\cite{lochner2016} and \cite{charnock2017} used simulated supernovae light curves from the \textit{Supernova Photometric Classification Challenge} (SPCC; \citealt{kessler2010}) to classify supernovae into three classes (Ia, Ib/c, II). \cite{muthukrishna2019} used a recurrent neural network to classify simulated ZTF light curves of various explosive transients, including supernovae. \cite{pasquet2019} used a convolutional neural network to classify supernovae light curves from multiple datasets (SPCC, simulated LSST, and Sloane Digital Sky Survey), capable of handling irregular sampling of light curves and a non-representative training set.
\cite{moller2020} developed a deep neural network approach to classify a set of simulated supernova light curves similar to the SPCC dataset, capable of classification on incomplete light curves. \cite{dauphin2020}, \cite{Hosseinzadeh2020}, and \cite{Villar2019} created classifiers trained on light curves of spectroscopically confirmed Pan-STARRS1 supernovae. \cite{takahashi2020} used a neural network to classify supernova light curves from the Hyper Suprime-Cam transient survey. In \citet{burhanudin2021}, we presented a recurrent neural network for classifying light curves from the Gravitational-wave Optical Transient Observer (GOTO) survey, capable of handling an imbalanced training dataset. In all the examples listed above, the overall classification performance is good for a number of classification tasks (binary Ia/non-Ia or a multi-class problem into the different supernova subtypes), achieving accuracies of $\gtrsim 85\%$.

The aforementioned studies all focus on classifying supernova light curves that have been obtained from a single survey or simulated to resemble the light curves of a particular survey. \cite{pruzhinskaya2019} used supernova light curves from multiple surveys, obtained from the Open Supernova Catalog \citep{guillochon2017} to develop an anomaly detection algorithm, capable of identifying rare supernovae classes and non-supernovae objects within the dataset. The challenge in working with light curves from different surveys is dealing with differences in how the photometry is calibrated, and the different filters used when making observations. By having a classifier that is agnostic to differences across different surveys, it allows the use of more available survey data so that the training of classifiers is not limited to the size of a sample obtained from just a single survey. We expand on the literature by applying a deep learning approach to classification on a heterogeneous dataset of supernova light curves, combined from multiple surveys. Using light curves from multiple surveys that use different filters also allows access to a wider wavelength coverage in broadband photometric observations.

One approach to working with a heterogeneous dataset is to find a way to standardise the data, so that they are represented in a more uniform manner. \citet{BooneAvocado2019} used a Gaussian process to model simulated LSST light curves from the Photometric LSST Astronomical Time Series Classification Challenge dataset (PLAsTiCC; \citealt{PLAsTiCC2018}), by interpolating in both time and wavelength. Gaussian processes have also been used to generate a two-dimensional representation of supernovae light curves \citep{qu2021, qu2021earlytime}, which are then used as inputs to a convolutional neural network for classification. By using a Gaussian process to interpolate light curves in time and wavelength, it is possible to create a uniform representation of light curves that consist of observations made across different filters. 

A challenge in creating classifiers for new surveys is the lack of a labelled training set with which to train a model. Many machine learning and deep learning classification methods assume that the training and test data come from the same distribution and share a common feature space. When the distribution changes, the models need to be retrained with a new labelled training set. In most cases, creating a new labelled training set to account for the change in distribution can be extremely difficult. Transfer learning \citep{pan2010} is an approach that uses the knowledge gained in performing a task (e.g. classification) in one domain (e.g. data from one particular survey) to perform another task in a different domain (e.g. classification using data from a different survey).

In this paper, we use Gaussian processes to create a uniform representation of supernova light curves from multiple surveys obtained through the Open Supernova Catalog, and use convolutional neural networks to classify the supernovae into different types. We also investigate the use of transfer learning to classify light curves from the PLAsTiCC dataset, using domain knowledge derived from the task of classifying Open Supernova Catalog light curves.
In section \ref{sec:osc_data} we introduce data from the Open Supernova Catalog, and in section \ref{sec:gp_interpolation} we present a two-dimensional Gaussian process to generate a two-dimensional representation of supernova light curves. Section \ref{sec:convnet} introduces the convolutional neural network used for classification. We present the results of classifying Open Supernova Catalog data in section \ref{sec:osc_results}. We introduce transfer learning and the PLAsTiCC data in section \ref{sec:transfer_learning_plasticc}, and results on classifying PLAsTiCC data in section \ref{sec:plasticc_results}. We provide a discussion of the work presented in this paper and conclude in section \ref{sec:discussion}.

\section{Open Supernova Catalog data}
\label{sec:osc_data}

Supernovae light curves and metadata (such as the supernova classification obtained via spectroscopy or through expert human inspection, any available spectroscopic data, and the R.A. and Dec) were retrieved from the Open Supernova Catalog website \footnote{https://sne.space/download/}. The downloaded data consisted of supernovae listed on the Open Supernova Catalog discovered up to the end of 2019, totalling $80,914$ objects. All objects that were labelled as `Candidate' or other non supernovae classes were discarded. Only objects that had been labelled as type Ia, Ibc, or II (including all sub-classifications within those types) were kept.

\subsection{Standardising magnitudes and filters}
\label{subsec:std_filters}

\begin{figure}
     \centering
     \includegraphics[width=0.47\textwidth]{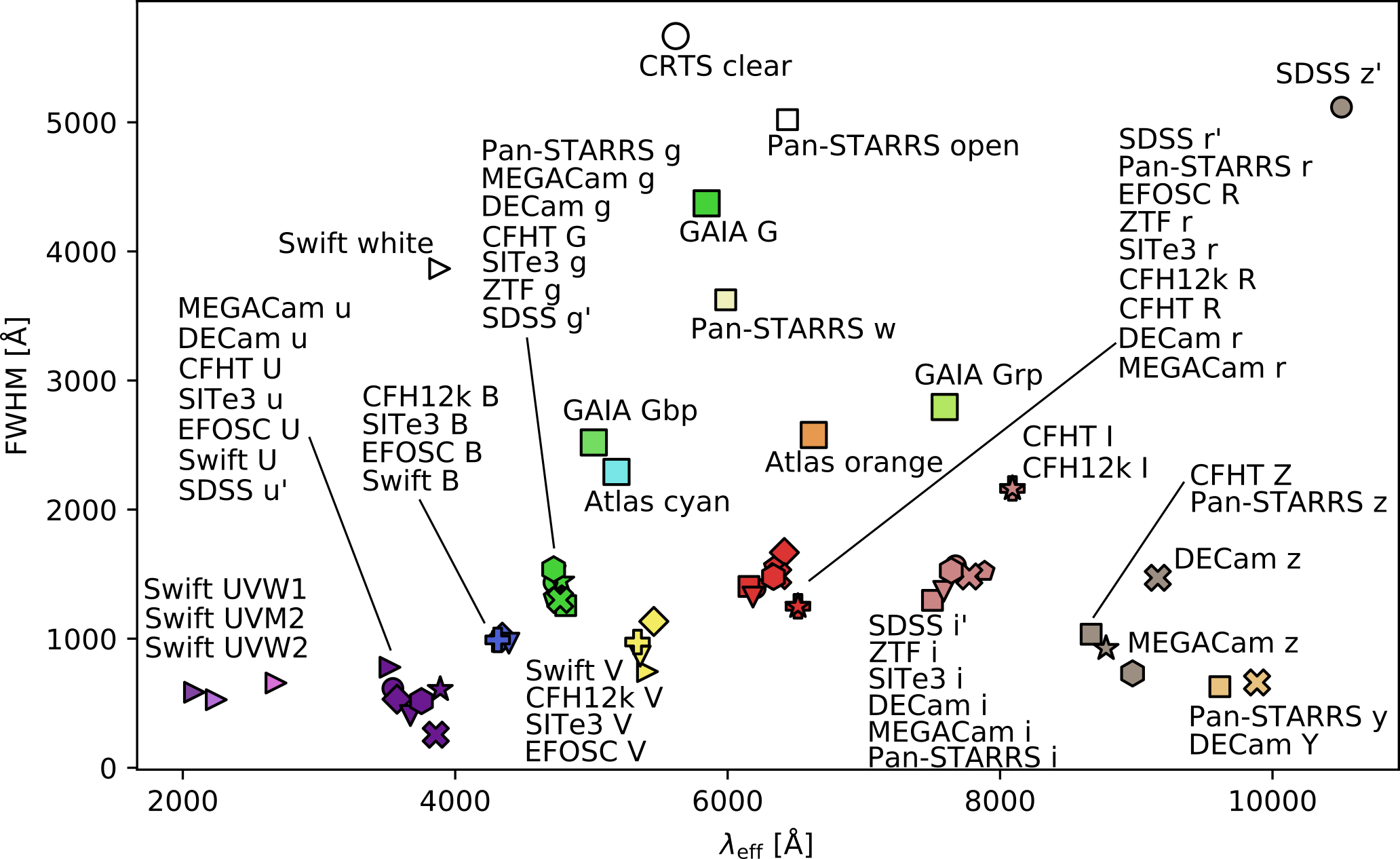}
     \caption{A plot of full width at half-maximum (FWHM) against the effective wavelength $\lambda_{\mathrm{eff}}$ of the filters available in the Open Supernova Catalog dataset. Both $\lambda_{\mathrm{eff}}$ and FWHM are given in Angstroms.}
     \label{fig:instrument_filters}
\end{figure}

The light curves in the Open Supernova Catalog dataset consist of observations that have been made with a variety of instruments across a number of different telescopes. In Figure \ref{fig:instrument_filters}, we plot the full width at half-maximum against the effective wavelength for the filters used in the dataset. The values for the effective wavelength and full width at half-maximum for the filters in Figure \ref{fig:instrument_filters} were obtained from the Spanish Virtual Observatory (SVO) Filter Profile Service \citep{svo2020}\footnote{http://svo2.cab.inta-csic.es/theory/fps/}.

The dataset also contains photometry generated using models or simulations and we discard these and only keep real photometric observations. Where available we use the \texttt{system} column in the photometry data to identify which magnitude system the data is calibrated to, otherwise the \texttt{instrument} column is used to identify which instrument was used to make the observation. The next step is then to convert all the magnitudes so that they are in the same magnitude system. The majority of magnitudes in the dataset are given in the AB magnitude system, so we convert the rest of the magnitudes into AB magnitudes. Magnitudes given in systems other than AB were converted using Tables \ref{tab:swift_AB}, \ref{tab:vega_AB}, and \ref{tab:CSP_AB} listed in the appendix. After converting all magnitudes into the AB system, the magnitudes were then converted into flux (in units of $\mathrm{erg}~\mathrm{s}^{-1}~\mathrm{Hz}^{-1}~\mathrm{cm}^{-2}$) using:

\begin{equation}
    f_\nu = 10^{-(m_{\mathrm{AB}}+48.60)/2.5}
\end{equation}
where $f_{\nu}$ is the monochromatic flux and $m_{\mathrm{AB}}$ is the AB magnitude.

\subsection{Light curve trimming}
\label{subsec:lc_trimming}

\begin{figure*}
    \centering
    \includegraphics[width=0.9\textwidth]{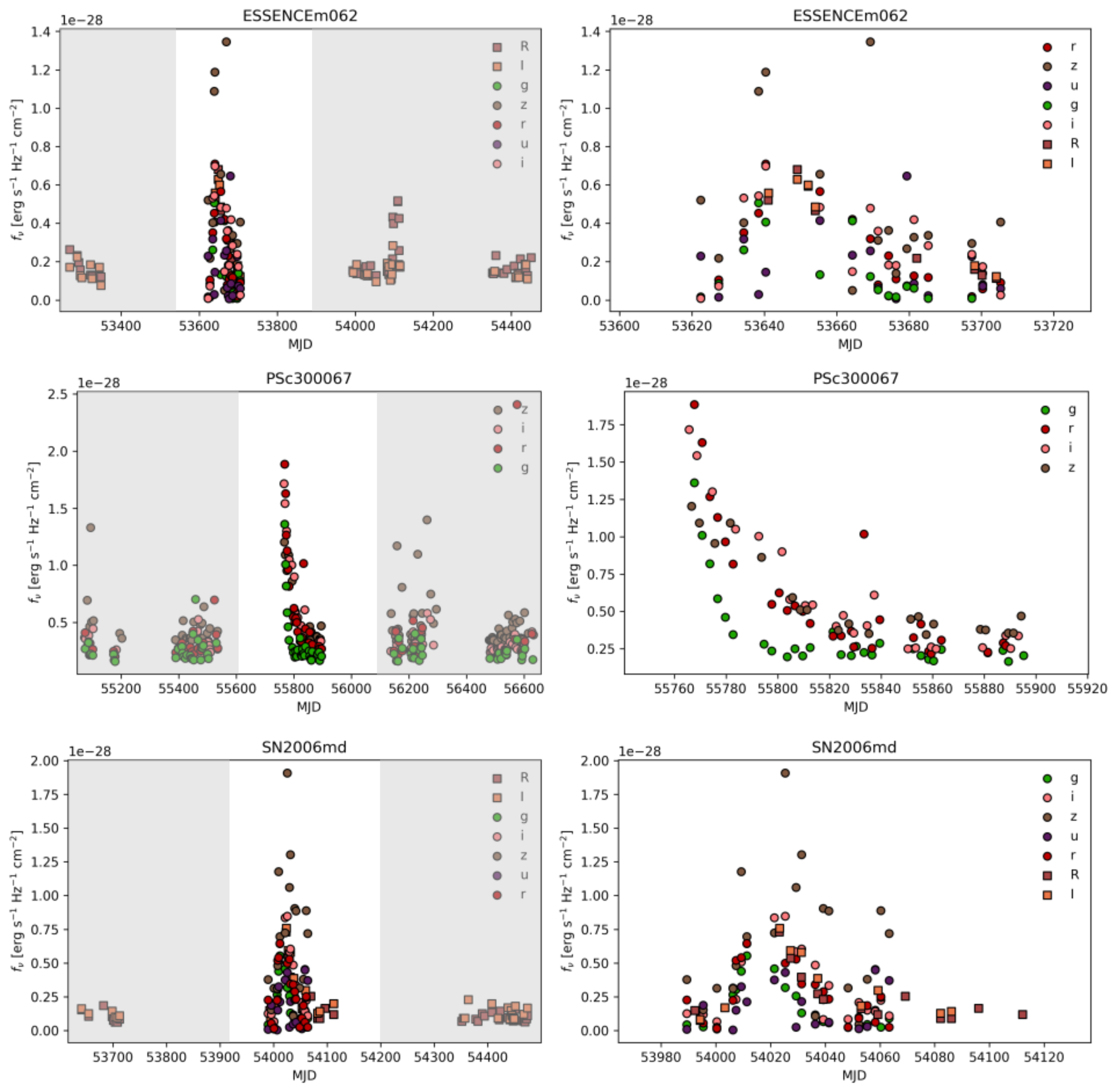}
    \caption{Light curves from the Open Supernova Catalog, before (left) and after trimming (right). The shaded regions on the left indicate parts of the light curves that were discarded.}
    \label{fig:trimmed_example}
\end{figure*}

Some light curves in the dataset span periods of up to multiple years, including seasonal gaps and periods where the only photometry available is actually of the host galaxy without the supernova. To shorten these longer light curves, so that we just consider the supernova, the steps listed below are taken:

\begin{itemize}
    \item Long light curves (longer than 300 days of observations) were split into shorter light curve chunks if there is a gap in observations longer than 60 days
    \item Compare the standard deviation in magnitudes of each light curve chunk $\sigma_{\mathrm{chunk}}$ to the standard deviation of the whole light curve $\sigma_{\mathrm{lc}}$
    \item If $\sigma_{\mathrm{chunk}} < \sigma_{\mathrm{lc}}$, then that portion of the light curve is discarded.  
\end{itemize}

Figure \ref{fig:trimmed_example} shows some example trimmed light curves. We find that this method is good at isolating the rise, peak, and decline of supernovae in the Open Supernova Catalog dataset.

\subsection{Selection cuts}
\label{subsec:cuts}

To create the final dataset, the following selection cuts were made:
\begin{itemize}
    \item Total number of observations in the light curve is $\geq6$
    \item At least two or more filters used
    \item The average number of observations per filter is $\geq2$
    \item The length of observations spans at least 20 or more days
\end{itemize}
\noindent These cuts were made to ensure that the light curves had good coverage across multiple wavelengths and in time so that there was enough information in each light curve to allow a model to learn to differentiate between the different classes. Good quality light curves are also required to ensure the light curve has enough data points to provide a good fit with Gaussian processes.

After the cuts, the number of remaining objects in our sample is 6330 of the total. The dataset is split into 60\% for training (3796 objects), 15\% for validation (951), and 25\% for testing (1583). The data is split into three classes for the classification task: Ia, Ibc, and II. The proportion of each class in the training, validation and test set is the same. This is done to ensure that there are a sufficient number of samples from each class in the validation and test set to evaluate classification performance across all classes, including those with fewer samples.

Table \ref{tab:osc_data_breakdown} summarises how the dataset is partitioned. Type Ia and II supernovae make up the majority of objects in the dataset, with only a small proportion of type Ibc supernovae in the data. This dataset presents a class imbalance problem, where one class contains fewer examples compared to other classes. Learning from an imbalanced dataset can be difficult, since conventional algorithms assume an even distribution of classes within the data set. Classifiers will tend to misclassify examples from the minority class, and will be optimised to perform well on classifying examples from the majority class (see \citealt{burhanudin2021}).

\begin{table}
    \centering
    \begin{tabular}{c|cccc}
        \toprule
         \textbf{Type} & \textbf{Training} & \textbf{Validation} & \textbf{Test} & \textbf{All data}\\
         \hline
         Ia & 2145 (56.5\%) & 542 (57.0\%) & 883 (55.8\%) & 3570 (56.4\%)\\
         II & 1563 (41.2\%) & 385 (40.5\%) & 657 (41.5\%) & 2605 (41.2\%)\\
         Ibc & 88 (2.3\%) & 24 (2.5\%) & 43 (2.7\%) & 155 (2.4\%)\\
         \hline
         Total: & 3796 & 951 & 1583 & 6330\\
         \bottomrule
    \end{tabular}
    \caption{A breakdown of how the final Open Supernova Catalog dataset (after light curve trimming and selection cuts) is divided for training, validation, and testing, along with the class distribution of the three supernova classes.}
    \label{tab:osc_data_breakdown}
\end{table}

\section{Gaussian processes for interpolation in time and wavelength}
\label{sec:gp_interpolation}

\subsection{Gaussian Processes}
\label{subsec:5_GPs}

A Gaussian process is a generalisation of the Gaussian probability distribution (which describe random variables) and can be thought of as a distribution over functions. Gaussian process regression attempts to find a function $f(x)$ given a number of observed points $y(x)$ that determines the value $y(x')$ for unobserved independent variables $x'$ (over a finite interval of $x'$ values) by drawing from a distribution of functions. The distribution of functions is determined by selecting a covariance function (also referred to as \textit{kernels}), which specifies the covariance between pairs of random variables. Covariance functions have adjustable hyperparameters, which determine the form of the Gaussian process prediction for $f(x)$. For a detailed discussion on Gaussian processes see \cite{rasmussen2005}.

\subsection{Two-dimensional Gaussian process regression}
\label{subsec:2d_gp}

In order to create a uniform representation of light curves in different filters, we follow the approach used in \citet{qu2021} and \citet{BooneAvocado2019}, and use two-dimensional Gaussian process regression to interpolate the light curves in wavelength and time. We model the light curves to create a two-dimensional image (referred to as a `flux heatmap' in \citealt{qu2021} and \citealt{qu2021earlytime}) where the flux is given as a function of time $t$ and wavelength $\lambda$.

We label each flux measurement in all light curves in the dataset with the effective wavelength $\lambda_{\mathrm{eff}}$ of the filter in which it was observed. The values for $\lambda_{\mathrm{eff}}$ for each filter are listed in Table \ref{tab:osc_lambda_eff}, and are obtained from the SVO Filter Profile Service \citep{svo2020} and \citet{blanton2007}. Observations covering wavelengths from the Swift UVW2 filter (with $\lambda_{\mathrm{eff}} = 2085.73${\AA}) up to the Johnson-Cousins $J$ filter (with $\lambda_{\mathrm{eff}} = 12355.0${\AA}) were used. These filters were used as the vast majority of observations in the dataset were made using filters within this wavelength range. All flux values of each light curve are associated with a time measurement $t$ (time of observations) and a wavelength value $\lambda_{\mathrm{eff}}$, the effective wavelength of the filter used to make the observation. We scaled the time so that the time of the first observations is $t=0$.

\begin{table}
    \centering
    \begin{tabular}{cc}
        \toprule
        Filter & $\lambda_{\mathrm{eff}}$ ({\AA})  \\
        \hline
        UVW2 & 2085.73\\
        UVW1 & 2684.14\\
        UVM2 & 2245.78\\
        $U$ & 3751.0\\
        $B$ & 4344.0\\
        $V$ & 5456.0\\
        $R$ & 6442.0\\
        $I$ & 7994.0\\
        $J$ & 12355.0\\
        $u$ & 3546.0 \\
        $g$ & 4670.0\\
        $r$ & 6156.0\\
        $i$ & 7472.0\\
        $z$ & 8917.0\\
        $y$ & 10305.0\\
        \bottomrule
    \end{tabular}
    \caption{The effective wavelengths $\lambda_{\mathrm{eff}}$ of the filters used to create flux heatmaps from the light curves in our sample dervied from the Open Supernova Catalog.}  
    \label{tab:osc_lambda_eff}
\end{table}

As in \citet{qu2021} and \citet{BooneAvocado2019}, we use the Mat\'ern 3/2 covariance function in our two-dimensional Gaussian process, with a fixed characteristic length scale in wavelength of $2567.32${\AA}, which is obtained by dividing the wavelength range covered by all the filters in Table \ref{tab:osc_lambda_eff} by 4. We note that this value is arbitrary, and that other values of fixed characteristic length scale could be used. \cite{BooneAvocado2019} find that their analysis on using Gaussian processes to model light curves is not sensitive to the choice of the length scale in wavelength, so we do not investigate this choice further. We leave the time length scale as a trainable parameter. The Mat\'ern 3/2 kernel has the form:

\begin{equation}
    k(r) = \sigma^2(1 + \sqrt{3r})\exp(-\sqrt{3r})
\end{equation}
where $\sigma^2$ is the variance parameter, which is left as a trainable parameter, and $r$ is the Euclidean distance between two input points $x_1$ and $x_2$, scaled by a length scale parameter $l$ (which we leave as a fixed constant):

\begin{equation}
    r = \frac{x_1 - x_2}{l^2}.
\end{equation}
The kernel used for the two-dimensional Gaussian process regression to model the light curves in wavelength and time is:

\begin{equation}
    k_{\mathrm{2D}} = \sigma^2 k_{\lambda}(r_\lambda) k_{t}(r_t)
\end{equation}
where $r_\lambda$ is the Euclidean distance between the wavelength input points, scaled by the fixed wavelength length scale parameter, and $r_t$ is the Euclidean distance between the time input points, scaled by the time length scale parameter.

\begin{figure*}
    \centering
    \includegraphics[width=0.72\textwidth]{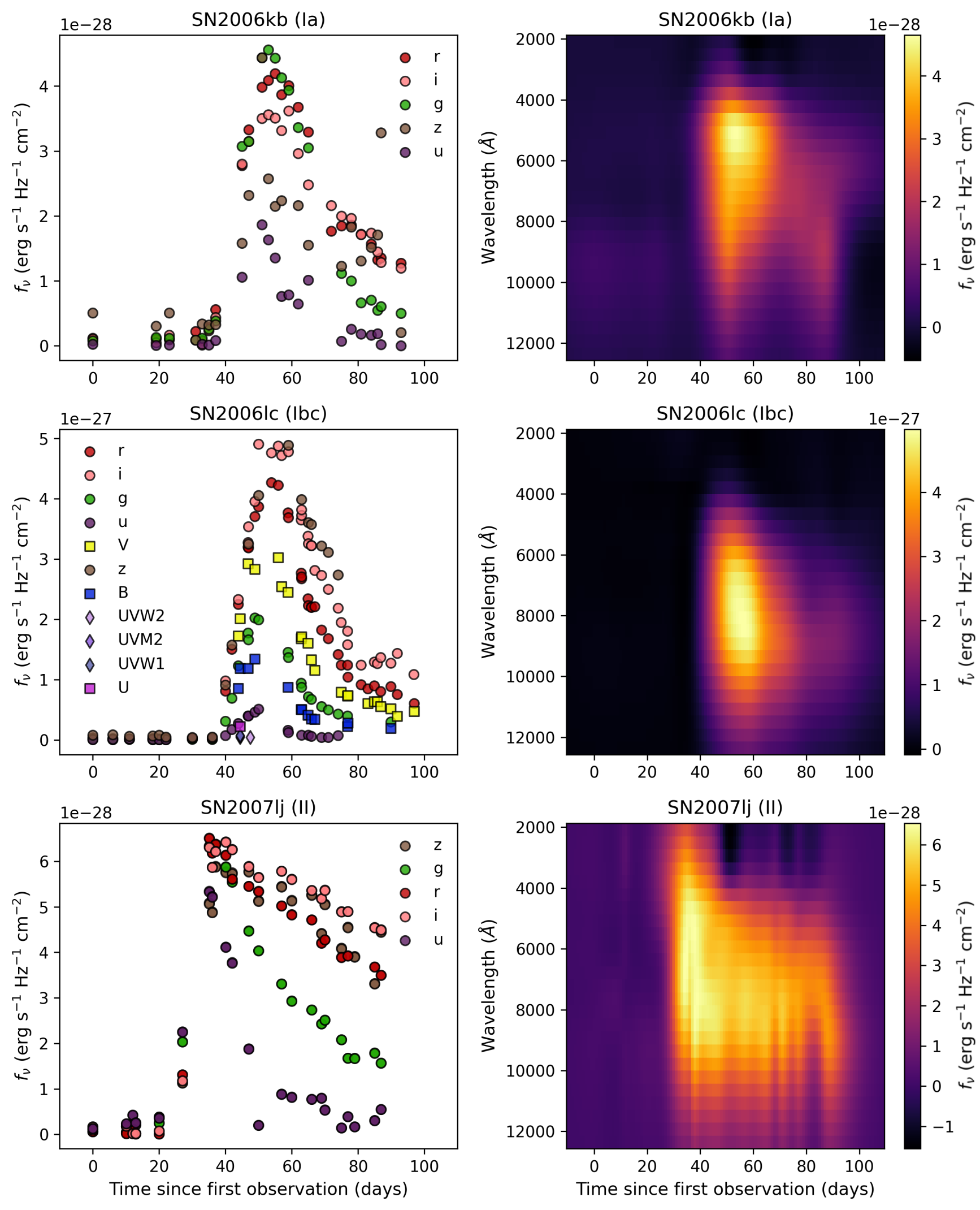}
    \caption{Examples of light curves (left) of SN2006kb (type Ia), SN2006lc (type Ibc), and SN2007lj (type II) and the corresponding flux heatmaps (right) generated using a two-dimensional Gaussian process. The light curves are plotted as flux $f_{\nu}$ converted from AB magnitudes in each filter against time. The heatmaps show flux (brighter pixels indicating higher flux values) as a function of time (in days) and wavelength (in \AA).}
    \label{fig:2d_gp_examples}
\end{figure*}


The two-dimensional Gaussian process is trained on each light curve in the dataset, and then used to predict flux measurements on a time-wavelength grid. The wavelength dimension in the grid runs from $2085.73${\AA} to $12355.0${\AA} divided into 25 bins resulting in a wavelength interval of $410.77${\AA}, and the time dimension runs from 10 days before and 110 days after the first observation with an interval of 1 day. In Section \ref{subsec:plasticc_heatmap} we use this approach to generate flux heatmaps for 397,990 PLAsTiCC supernova light curves, so the choice of dimensions is a practical one. The resulting flux heatmap image has dimensions of $120 \times 25$ pixels, where each pixel represents a flux measurement. Figure \ref{fig:2d_gp_examples} shows example flux heatmaps. The flux heatmaps are used as input for a convolutional neural network for classification. We use the \texttt{GPFlow} Python package \citep{GPflow2017} to perform Gaussian process regression. The time taken to generate a flux heatmap from a light curve with a Gaussian process on a 4-core CPU is approximately 3 seconds.

\subsection{Using two-dimensional Gaussian processes to infer spectra from light curves}
\label{subsec:2d_gp_for_spectra}

\begin{figure*}
    \centering
    \includegraphics[width=0.85\textwidth]{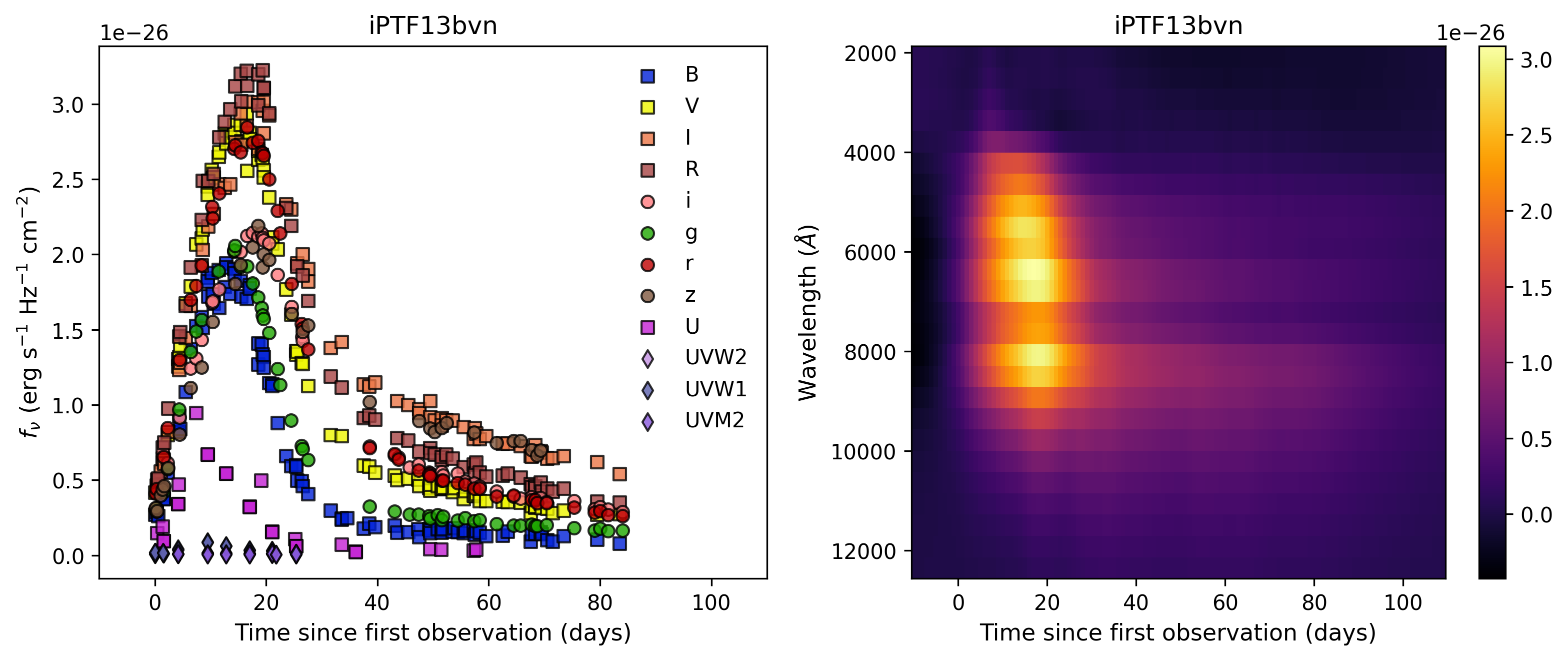}
    \caption{The light curve of the type Ib supernova iPTF13bvn (left) and its flux heatmap generated from the light curve (right).}
    \label{fig:iPTF13bvn_gp}
\end{figure*}

The two-dimensional Gaussian process regression can be used as a method to infer supernova spectra from their light curves. We select iPTF13bvn from the Open Supernova Catalog dataset as an example. This is a type Ib supernova that has good photometric coverage in time and across multiple filters. Figure \ref{fig:iPTF13bvn_gp} shows the light curve and the corresponding flux heatmap generated with a two-dimensional Gaussian process. 

\begin{figure*}
    \centering
    \includegraphics[width=0.7\textwidth]{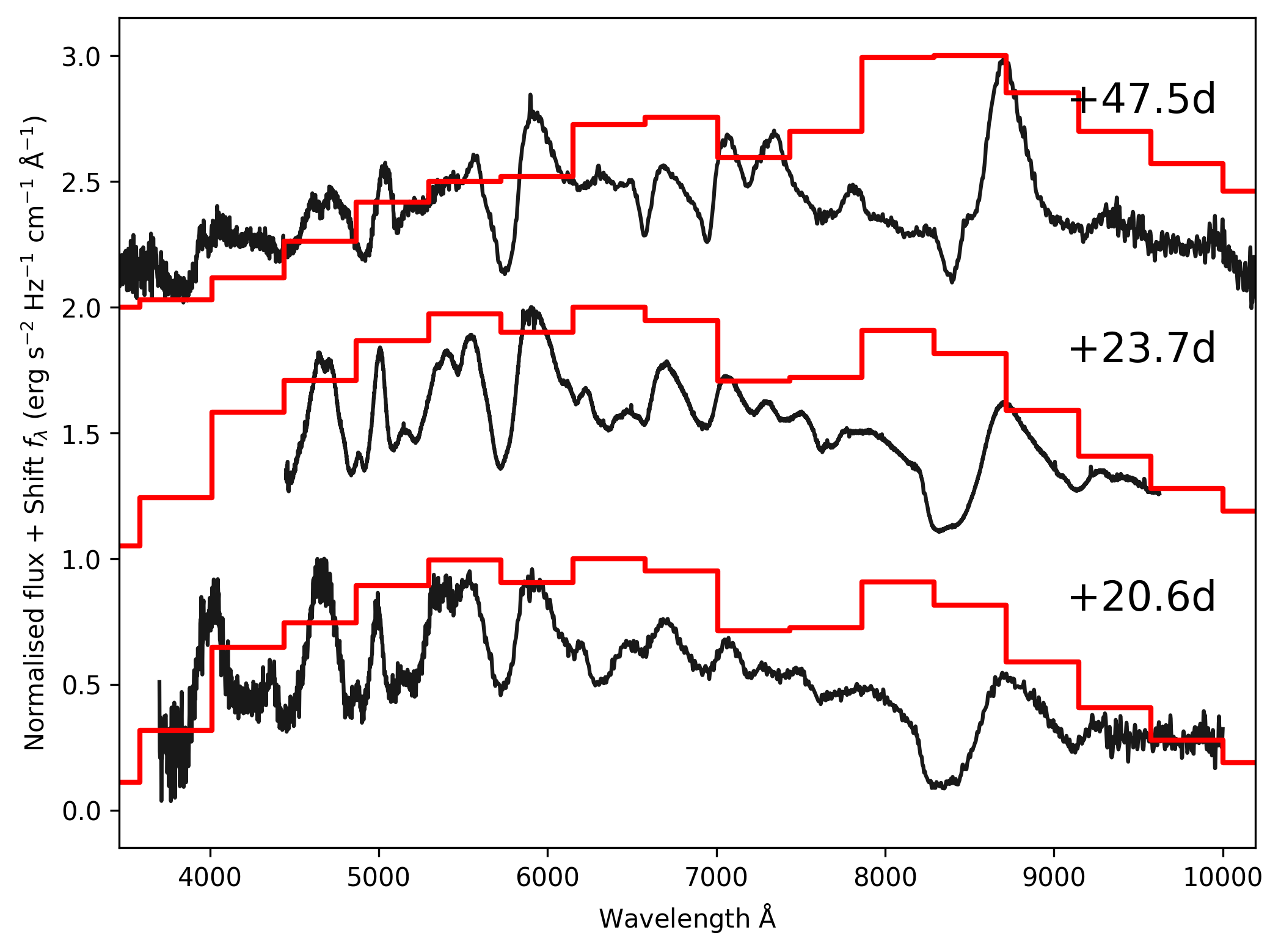}
    \caption{The spectra of iPTF13bvn are shown in black, and the spectra obtained from the flux heatmap are shown in red. The time of the spectra is given as days from the time of the first observation of the light curve. The spectra have been normalised (using the maximum value for each individual spectra) and shifted for clarity.}
    \label{fig:iPTF13bvn_spectra}
\end{figure*}

We examine three spectra for iPTF13bvn, made available through the Open Supernova Catalog \citep{guillochon2017, shivvers2019}. To obtain the `simulated' spectra from the flux heatmap, we take a single column at the time the spectra were taken, giving a vector that measures flux as a function of wavelength. The time of observation of the spectra is scaled to the time of first observation in the light curve, so it is given as the number of days since the first light curve observation. The real spectra for iPTF13bvn are taken at 20.6, 23.7, and 47.5 days after the first light curve observation, so the corresponding simulated spectra are obtained by taking columns from the flux heatmap at 20, 24, and 48 days after the first light curve observation. Figure \ref{fig:iPTF13bvn_spectra} compares the real spectra of iPTF13bvn to the simulated spectra obtained from the flux heatmap.

From Figure \ref{fig:iPTF13bvn_spectra}, it can be seen that the spectra generated from the flux heatmap correlate quite well with the real spectra of iPTF13bvn. In all three spectra, the heatmap generated spectra appear to trace the continuum shape. For the spectra obtained at 47.5 days, the heatmap generated spectrum correlates with the Ca II IR triplet emission feature at $\sim8700$\AA. Although there is a correlation, there is a poor match between the real spectrum and the heatmap generated spectrum which could be due to the width of the red filters (see Figure \ref{fig:instrument_filters}). Here, we have shown one example where the two-dimensional Gaussian process to create a flux heatmap can be used to generate low resolution spectra, provided there is good photometric coverage across multiple filters. In this example, photometric data for iPTF13bvn was obtained from three different sources across 12 different filters.


\section{Convolutional neural networks}
\label{sec:convnet}

\begin{figure}
    \centering
    \includegraphics[width=0.4\textwidth]{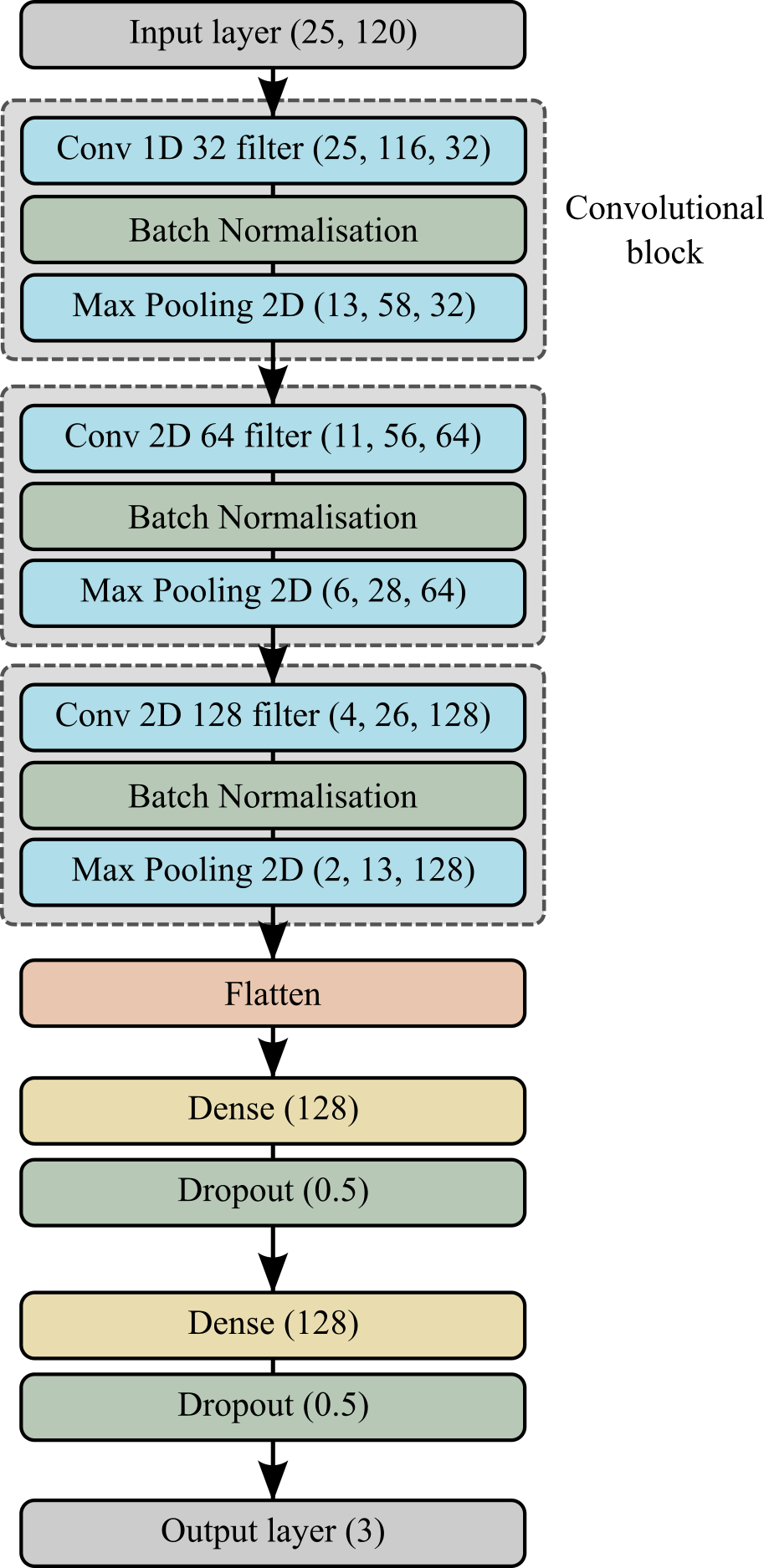}
    \caption{A diagram of the convolutional neural network used in this paper. The grey dashed box indicates the layers that make up a convolutional block. The dimensions of the output tensors in the layers in the convolutional blocks (Conv 1D, Conv 2D, Max Pooling 2D), number of neurons in the dense layers (Dense), and the dropout fractions (Dropout) are shown in parentheses.}
    \label{fig:cnn_diagram}
\end{figure}

\subsection{Model architecture}

Convolutional neural networks (CNNs; \citealt{lecun1989cnn}) are a class of neural networks that can process data with a grid-like structure (e.g. a two-dimensional grid of pixels in an image, or a sequence of measurements in time-series data where there may be one or more measurements at each time step). CNNs use convolution filters to identify spatial features in the data (such as corners or edges in an image). The output of a convolution applied to the input data is referred to as a `feature map'

We use a CNN to classify the flux heatmaps created from the Open Supernova Cataolog light curves (section \ref{subsec:2d_gp}) into three different classes: supernovae of types Ia, Ibc, and II. We build the CNN using the TensorFlow 2.0 package for Python \citep{tensorflow2016}\footnote{https://www.tensorflow.org/} with Keras \citep{chollet2015keras} for implementation of network layers.

The input to the CNN is a two-dimensional flux heatmap image of a supernova light curve, with dimensions $120 \times 25$ pixels where each pixel represents a flux density value. All flux heatmaps are normalised by dividing by the highest flux density value, so that the pixels in every heatmap have values between $0$ and $1$. We use a CNN with three convolutional blocks, followed by two fully-connected layers before the final output layer. Each convolutional block consists of a convolution layer, a batch normalisation layer, and a $2 \times 2$ max pooling layer. Figure \ref{fig:cnn_diagram} illustrates the model architecture used. For the convolutional and dense layers, the rectified linear unit (ReLU) activation function is used, and in the final output layer the softmax activation function is used to produce a list of probabilities that sum to unity. The probabilities returned by the model are scores that describe the level of `belongingness' to a class.


In the first convolutional block we apply a one-dimensional convolution (also called a temporal convolution) in the time dimension instead of a standard two-dimensional convolution. This is done since the flux heatmap is generated from a light curve which measures the brightness of a supernova over time, so  we attempt to extract temporal features in the first convolutional block. The output of each convolutional block has dimensions $(n_{\mathrm{rows}}, n_{\mathrm{columns}}, n_{\mathrm{filters}})$, where $n_{\mathrm{filters}}$ is a convolutional layer parameter and refers to the number of convolution filters used to generate feature maps. In the second and third convolutional blocks, a two-dimensional convolution is applied to the output of the preceding convolutional block. Table \ref{tab:model_conv} lists the series of convolutions and max pooling applied in the convolutional blocks, with the corresponding layer parameters and output dimensions at each stage.

\begin{table*}
    \centering
    \begin{tabular}{cccc}
        \toprule
         \textbf{Layer} & \textbf{Kernel/Pool size} & \textbf{Filters} & \textbf{Output dimension} \\
        \hline 
        Conv 1D & (5) & 32 & (25, 116, 32)\\
        BatchNorm & - & - & (25, 116, 32)\\
        MaxPool 2D & (2,2) & - & (13, 58, 32)\\
        \hline
        Conv 2D & (3,3) & 64 & (11, 56, 64)\\
        BatchNorm & - & - & (11, 56, 64)\\
        MaxPool 2D & (2,2) & - & (6, 28, 64)\\
        \hline
        Conv 2D & (3,3) & 128 & (4, 26, 128)\\
        BatchNorm & - & - & (4, 26, 128)\\
        MaxPool 2D & (2,2) & - & (2, 13, 128)\\
        Flatten & - & - & 3328\\
        \bottomrule
    \end{tabular}
    \caption{The layer parameters and output dimension for each layer in the convolutional blocks. For the convolutional layers, the kernel size is the shape of the convolutional window and filters sets the number of convolutional filters that are learnt during training. For the max pooling layers, the pool size sets the shape of the window over which to take the maximum. The number of strides is one for the convolutional layers and two for the max pooling layers. The flattening layer takes the multidimensional output of the convolutions and shapes it into a single dimensional output. }
    \label{tab:model_conv}
\end{table*}

The output of the last convolutional block is then flattened into a one-dimensional vector and then passed on to two fully-connected layers, each with dropout applied with the dropout fraction set to 0.5. We apply a $L_2$ regularization in the second fully-connected layer with a regularization parameter of 0.01, which is the default TensorFlow value. The final output layer is a fully-connected layer with the same number of neurons as the number of classes, which is three. In total, the CNN model has 536,003 trainable parameters.

\subsection{Model training}
\label{subsec:osc_training}

\begin{figure}
    \centering
    \includegraphics[width=0.45\textwidth]{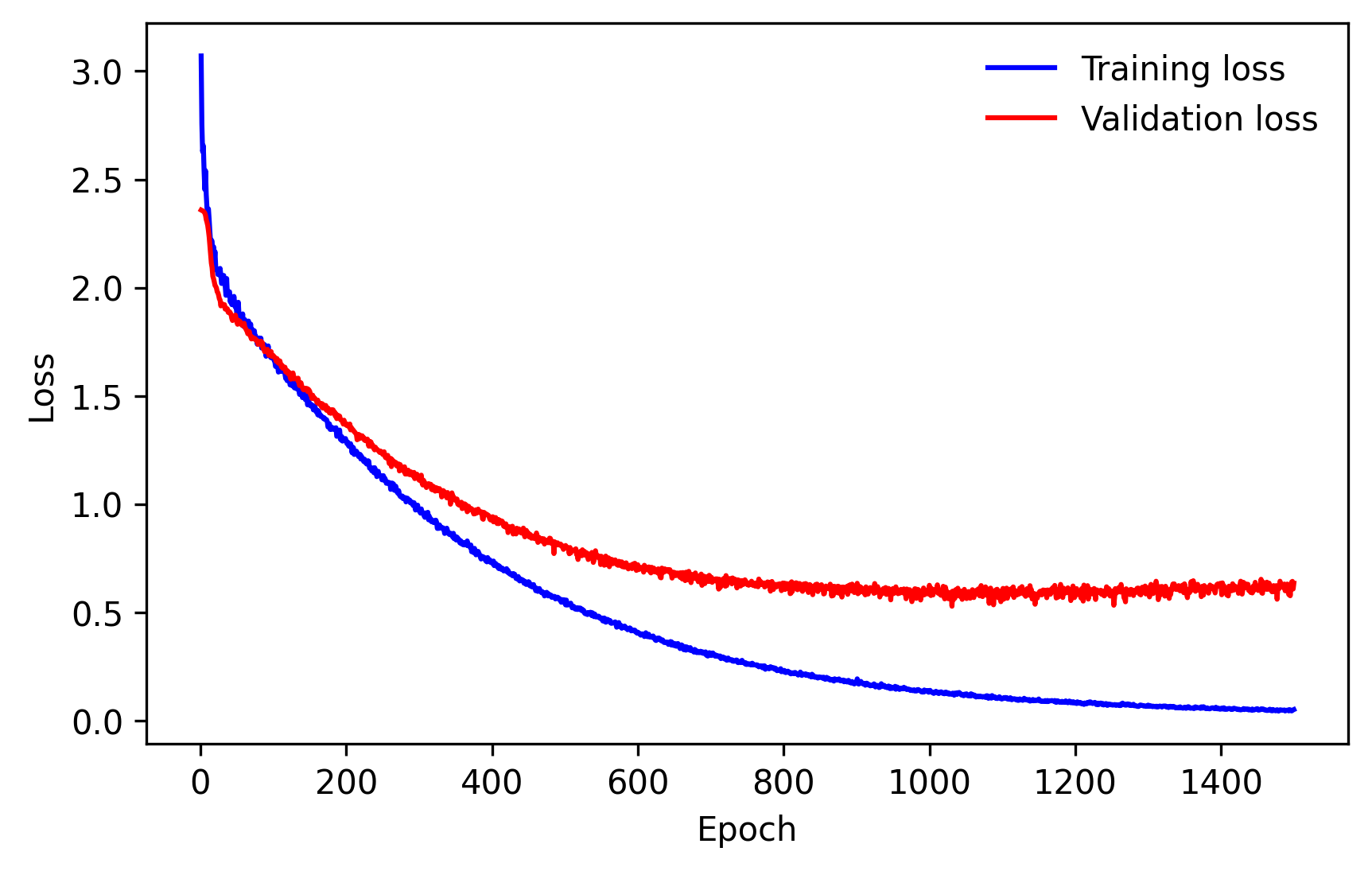}
    \caption{The training and validation loss for the CNN model trained on the Open Supernova Catalog data.}
    \label{fig:osc_training}
\end{figure}

The CNN model is trained on the flux heatmaps generated from the Open Supernova Catalog light curves with a learning rate of $1\times 10^{-5}$ for 1500 epochs with the Adam optimizer \citep{kingma2014}, using a batch size of 128. Figure \ref{fig:osc_training} shows how the training and validation loss evolve with training. Within 1500 epochs of training, both the training and validation loss begin to converge (i.e. stops improving). We use a cross entropy loss function, weighted to take into account the class imbalance present in the data. Given a multi-class problem with $N$ classes, the cross entropy loss (CE) for an example $i$ is:

\begin{equation}
\label{eq:crossentropy}
    \mathrm{CE} = 
    -\displaystyle\sum_{j=1}^{N}\delta_{ij}\alpha_{j}\log(p_{ij}),
\end{equation}
where $p_{ij}$ is the probability of example $i$ belonging to class $j$, $\alpha_j$ is the class weight for class $j$, and $\delta_{ij}$ is the Kronecker delta function. The loss for the entire data set is given by summing the loss of all examples. The class weight $\alpha_j$ for class $j$ is
\begin{equation}
    \alpha_j = \frac{1}{n} \times \frac{N}{N_j}
\end{equation}
where $n$ is the total number of classes, $N$ is the total number of samples in the dataset, and $N_j$ is the number of samples in class $j$. The class weights are obtained using samples in the training set. 

The model is trained on an NVIDIA Quadro P2200 graphics processing unit with 1280 cores and 5GB of memory, which takes 4 seconds per epoch for a total time of $\sim 100$ minutes to train the model.

\section{Results on classifying Open Supernova Catalog data}
\label{sec:osc_results}

\begin{figure}
    \centering
    \includegraphics[width=0.42\textwidth]{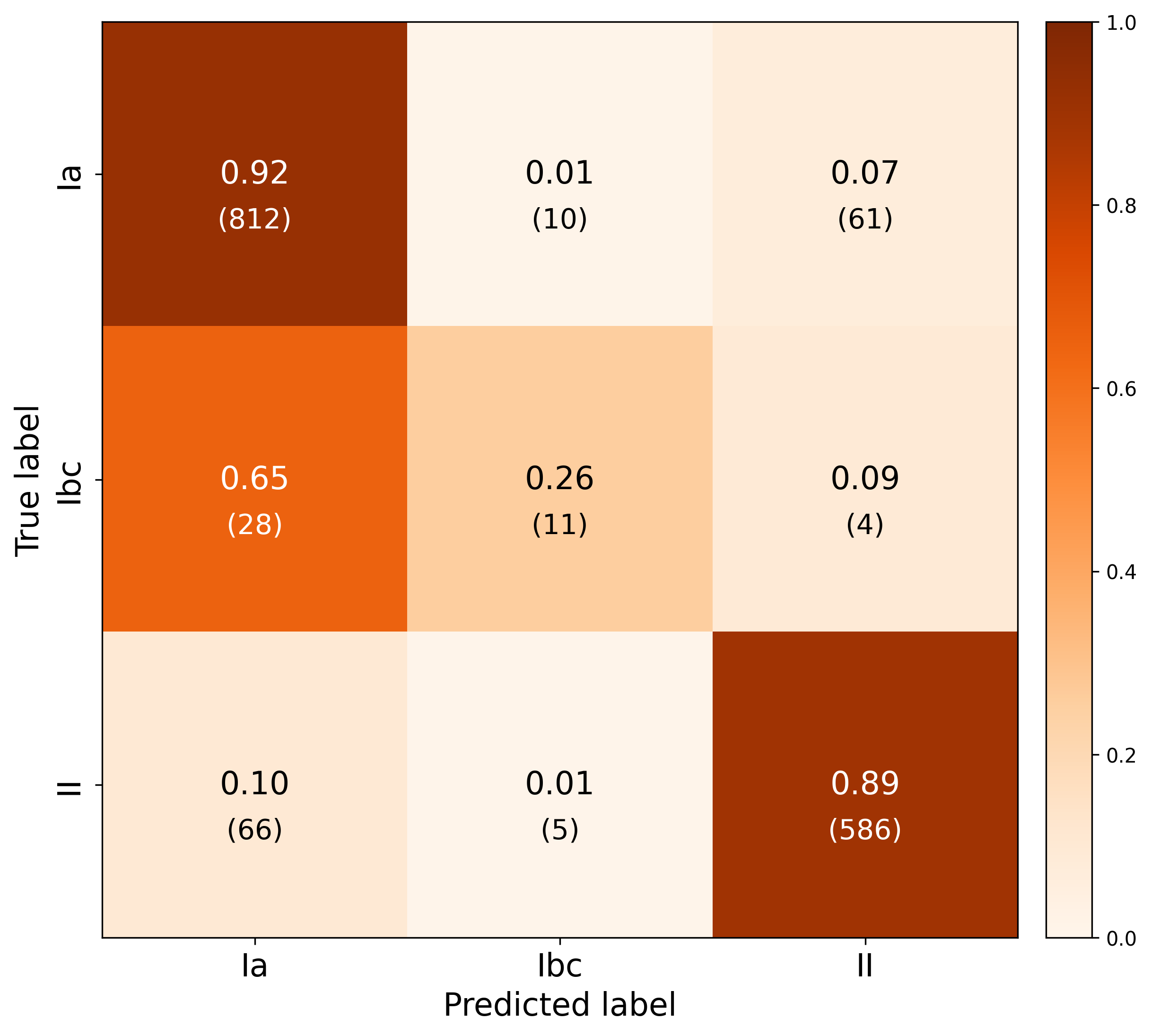}
    \caption{Confusion matrix for the test set of flux heatmaps generated from the Open Supernova Catalog light curves. The y-axis shows the true class label, and the x-axis shows the class label predicted by the model. Entries along the diagonal represent where the predicted label matches the true label, and the off-diagonal entries show where misclassifications occur. Reading along the rows, the fractional values show how samples from a class have been classified, with the absolute numbers below in parentheses.}
    \label{fig:osc_test_cm}
\end{figure}

Once the model has been trained, it is then used to make predictions on the test set. The test set consists of data that is kept apart from the training and validation sets, and used to evaluate how well the model is able to generalize on unseen data. On the test set, the model achieves an Area Under the Receiver Operating Characteristic curve (AUC) score of 0.859, and an $F_1$ score of 0.708. Figure \ref{fig:osc_test_cm} shows the confusion matrix for the model evaluated with the test set.

From the confusion matrix, the model shows good classification of Type Ia and II supernovae with 92\% (812) and 89\% (586) accuracy for each class, respectively. The performance for Type Ibc supernovae is poor, with the model only achieving 26\% (11) accuracy for that class and misclassifying 65\% (28) of Type Ibc supernovae as Type Ia. This may be due to the small number of samples of type Ibc supernovae in the dataset. The majority of Type Ibc supernovae are misclassified as Type Ia, and it is known that it can be challenging to differentiate between Type Ia and Type Ibc with only photometry \citep{lochner2016}. The misclassifications between Type Ia and Type II are quite low, with $\lesssim~10\%$ (61 for Type Ia and 66 for Type II) of each being misclassified as the other.

\section{Transfer learning on PLAsTiCC supernova light curves}
\label{sec:transfer_learning_plasticc}

\subsection{Overview}
\label{subsec:transfer_overview}

In the case of a classification task where there is a lack of labelled training data, the ability to transfer classification knowledge from one domain to a new one is useful. In astronomy, new surveys can experience the problem of a small or complete lack of a labelled training set since it can take time to accumulate enough sources and also label them (e.g. using spectroscopy or visual inspection of the photometry). In the following sections, we present the application of transfer learning to classify supernova light curves from the Photometric LSST Astronomical Time Series Classification Challenge (PLAsTiCC) dataset \citep{PLAsTiCC2018} by using classification knowledge derived from the Open Supernova Catalog light curves presented in the previous sections.

Transfer learning is defined as improving the learning of a target predictive function (e.g classification, mapping inputs to a class) in a target domain $\mathcal{D}_T$ using knowledge from a source domain $\mathcal{D}_S$ and source task $\mathcal{
T}_S$ \citep{pan2010}. In this case, the target domain is the PLAsTiCC dataset, the source domain is the Open Supernova Catalog dataset, the source task is classifying Open Supernova Catalog light curves into one of three classes (Ia, Ibc, II), and the target predictive function is classifying light curves from the PLAsTiCC dataset.

\subsection{The PLAsTiCC dataset}
\label{subsec:target_domain}

The Photometric LSST Astronomical Time Series Classification Challenge (PLAsTiCC) was launched in 2018 to challenge participants from the wider science community (open to not just astronomers but experts in other fields such as computer science) to develop classification algorithms or models to classify a large dataset of simulated LSST observations \citep{PLAsTiCC2018}.

The dataset consisted of over 3.5 million objects with a total of over 450 million observations, divided into a wide range of classes (supernovae of various types, variable objects, tidal disruption events, and more), each with light curves in six filters (LSST $ugrizy$) that include the fluxes and corresponding errors, with the time of observation. Approximately 8,000 objects were provided with labels that formed a mock `spectroscopically confirmed' training set. After the challenge was completed, an `unblinded' dataset was released with labels for all objects in the test set. We use the unblinded dataset in this paper. For each object, contextual information such as the R.A. and Dec, Galactic latitude and longitude, and host galaxy spectroscopic and photometric redshifts were available. 

The PLAsTiCC dataset presents its own unique set of challenges, such as the presence of `seasonal gaps' in the light curves where an object is not visible during the observing campaign, a wide distribution of class sizes (with some classes having only hundreds of examples vs. others having millions), and a training set that is not representative in redshift of the test set (to simulate a realistic training set obtained from a spectroscopically confirmed sample of nearby and brighter objects).

\subsection{The new classification problem}
\label{subsec:target_task}

Transfer learning can be used to borrow classification knowledge from one task in one domain to another task in another domain. As defined above, the domains are the two different datasets. We also define a new classification task for the PLAsTiCC dataset, which is different to the classification task presented in section \ref{sec:osc_data}. We select only supernovae from the PLAsTiCC dataset, and define six classifications based on the PLAsTiCC defined classifications in \cite{PLAsTiCC2018}: types Ia, Iax, Ia-91bg, Ibc, II, SLSN-I. The classifications now divide type Ia supernovae into three sub-classes, and and also include a new class, type I superluminous-supernovae (SLSN-I). 

\subsection{Data selection}
\label{subsec:data_selection}

Light curves in the PLAsTiCC dataset span the duration of the observing campaign and feature seasonal gaps when an object is not observable. We use only photometry obtained from the image-subtraction pipeline (using the flag \texttt{detected\_bool} = 1), which removes the seasonal gaps and produces light curves covering the period of supernova rise and decline. We also select only observations from up to 20 days before and 100 days after the peak (which is taken as the maximum flux measurement in any filter).

After applying the selection cuts, the final dataset consists of $397,990$ objects with $2,398$ labelled objects remaining from the original training set. For the transfer learning process, we use two training sets and compare their performance. The first is the original training set, and the second is an augmented training set which is obtained by randomly sampling $3\%$ of the test set added to the original training set. No stratification is used when sampling the test set, so the proportion of the six classes is unchanged, still presenting a class imbalance problem. The test set with the $3\%$ removed for creation of the augmented training set is used as the test set for both the original training set and the augmented training set, so that the models trained on the two training sets are evaluated on the same test set. In both training sets, 10\% is used for validation. Table \ref{tab:plasticc_data_breakdown} shows the breakdown of the PLAsTiCC dataset. The original PLAsTiCC training set contains fewer samples than the Open Supernova Catalog training set. This transfer learning approach emulates using classification knowledge from another domain after a small labelled training set has been obtained for a new survey.

\begin{table*}
    \centering
    \begin{tabular}{c|ccc|c}
        \toprule
         \textbf{Type} & \textbf{Original} & \textbf{Augmented} & \textbf{Test} & \textbf{All data}\\
         \hline
         Ia & 1,136 (47.4\%) & 7,168 (50.5\%) & 197,884 (51.9\%) & 206,188 (51.8\%)\\
         Iax & 115 (4.8\%) & 341 (2.4\%) & 6,730 (1.8\%) & 7,186 (1.8\%)\\
         Ia-91bg & 78 (3.3\%) & 197 (1.4\%) & 4,151 (1.1\%) & 4,426 (1.1\%)\\
         Ibc & 259 (10.8\%) & 1,082 (7.6\%) & 26,271 (6.9\%) & 27,612 (6.9\%)\\
         II & 670 (27.9\%) & 4,735 (33.4\%) & 128,958 (33.8\%) & 134,363 (33.8\%)\\
         SLSN-I & 140 (5.8\%) & 671 (4.1\%) & 17,404 (4.6\%) & 18,215 (4.6\%)\\
         \hline
         Total: & 2,398 & 14,194 & 381,398 & 397,990\\
         \bottomrule
    \end{tabular}
    \caption{Breakdown of the PLAsTiCC dataset by type, for the original, augmented, and test sets.}
    \label{tab:plasticc_data_breakdown}
\end{table*}

\subsection{Creating heatmaps}
\label{subsec:plasticc_heatmap}

We follow the same steps outlined in section \ref{sec:gp_interpolation}, and use a two-dimensional Gaussian process to generate heatmaps from the PLAsTiCC supernova light curves. The time of observation was scaled so that the time of the first observation is $t=0$. Each flux measurement in a light curve is labelled with the time it was observed, and the effective wavelength $\lambda_{\mathrm{eff}}$ of LSST filter it was observed in. Table \ref{tab:lsst_filters} lists the effective wavelengths for the LSST filters. 

\begin{table}
    \centering
    \begin{tabular}{cc}
    \toprule
    \textbf{Filter} & $\pmb{\lambda_{\mathrm{eff}}}$\textbf{(\AA)} \\
    \hline
    $u$ &  3751.36\\
    $g$ & 4741.64\\
    $r$ & 6173.23\\
    $i$ & 7501.62\\
    $z$ & 8679.19\\
    $y$ & 9711.53\\
    \bottomrule
    \end{tabular}
    \caption{The effective wavelength $\lambda_{\mathrm{eff}}$ of the LSST filters used to simulate observations in the PLAsTiCC dataset. The values were obtained from the SVO Filter Profile service \citep{svo2020}.}
    \label{tab:lsst_filters}
\end{table}

 \begin{figure}
     \centering
     \includegraphics[width=0.49\textwidth]{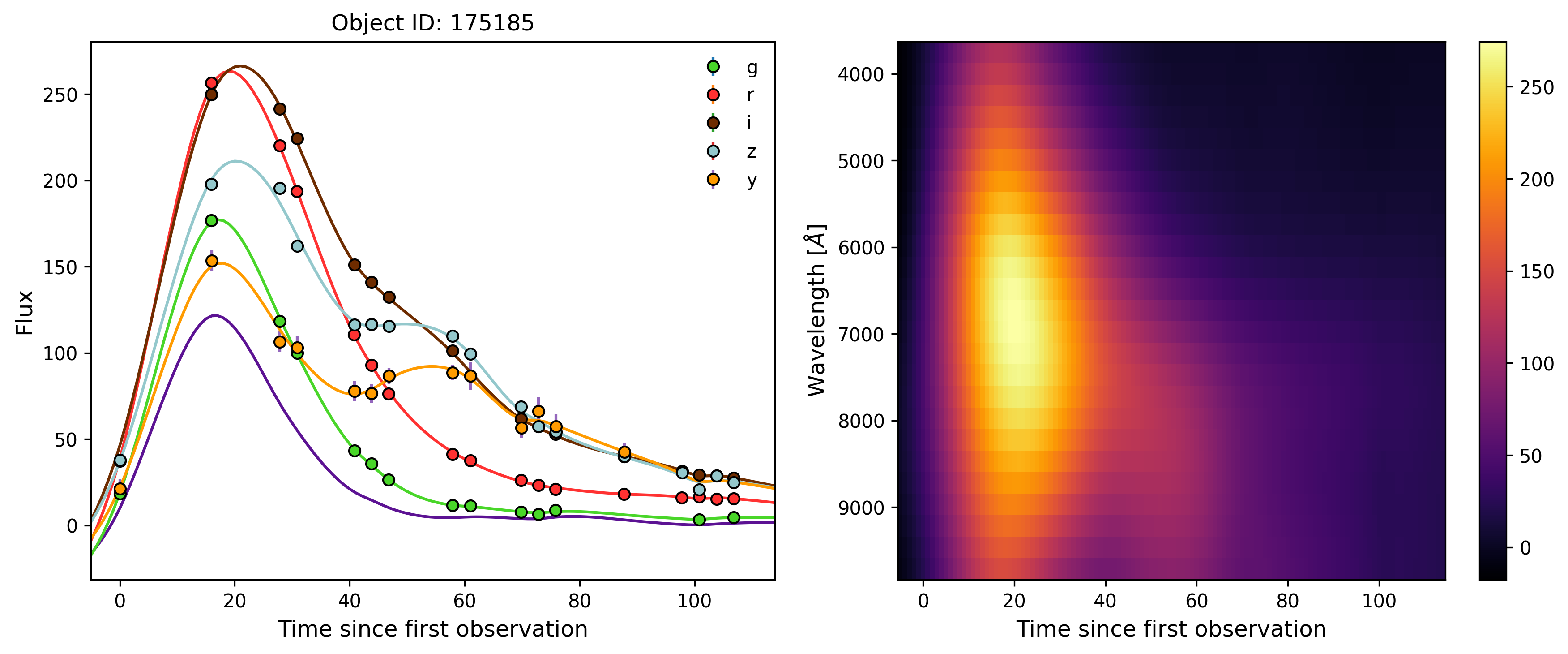}
     \caption{An example of a type Ia supernova light curve (left) and the flux heatmap generated from the light curve (right). The interpolated flux from the two-dimensional Gaussian process at the wavelength corresponding to the filter effective wavelength is also plotted for the light curve on the left.}
     \label{fig:plasticc_gp_example}
 \end{figure}

The light curves were then used to train a two-dimensional Gaussian process to create flux heatmaps. A fixed characteristic wavelength scale of $2980.09${\AA} was used, obtained by dividing the wavelength range coverage of the filters by two (to produce a similar value for the fixed wavelength scale in Section \ref{sec:gp_interpolation}). The time length scale and variance parameter were left as trainable parameters. The flux heatmaps were generated onto a grid, where $-5<t<115$ with a one-day interval and wavelength running from $3751.36${\AA} to $9711.53${\AA} , divided into 25 bins giving an interval of $238.81${\AA}. The flux heatmaps have dimensions of $120 \times 25$ pixels, where each pixel represents a flux value. Figure \ref{fig:plasticc_gp_example} shows an example light curve and the flux heatmap generated using the two-dimensional Gaussian process.

\subsection{Applying transfer learning to PLAsTiCC light curves}
\label{subsec:applying_tl_plasticc}

We compare two models on their classification performance on the PLAsTiCC dataset, one with transfer learning and without. In both cases we examine how including redshift information and using the augmented training set affects performance. We use the estimated host galaxy photometric redshift value in the PLAsTiCC data listed in the \texttt{hostgal\_photoz} column. We use the same CNN model architecture presented in Section \ref{sec:convnet}, but change the output layer to have six neurons (for the six classes in the PLAsTiCC classification task). For the models that include redshift information, we append the redshift value to the flattened output of the last convolutional block.

Transfer learning is implemented by setting the parameters of the convolutional block as non-trainable parameters, a method known as `freezing' layers in a neural network. The parameters in the convolutional blocks are fixed, and the model only changes the parameters in the dense layers during training. The idea is that the `knowledge' of extracting salient features from the heatmaps developed in the convolutional blocks of the model trained on the Open Supernova Catalog dataset is used to extract features from heatmaps in the PLAsTiCC dataset. Since the only trainable parameters are those in the dense layers, the model is then just tasked with learning the feature-class relationship to group the data into different classes using the features extracted from the heatmaps. 

For the models without transfer learning, all parameters are left as trainable parameters. In this case, the model has to learn to extract features from heatmaps in the convolutional blocks as well as the feature-class relationship in the dense layers to classify the heatmaps into the six classes. Since the models used in transfer learning have fewer trainable parameters, the time needed to train them is less than the time needed to train the models without transfer learning. The models were trained on a NVIDIA Quadro P2200 graphics processing unit with 1280 cores and 5GB of memory, and the models with transfer learning required 0.29s per epoch of training, while the model without transfer learning required 0.51s per epoch. With transfer learning, the models could be trained $\sim 56\%$ faster. All models are trained for 500 epochs with a learning rate of $1\times10^{-4}$. The training history for the transfer learning models are shown in the appendix \ref{app:transfer_learning}.

\section{Classifying PLAsTiCC light curves with transfer learning}
\label{sec:plasticc_results}

\subsection{Models without transfer learning}
\label{subsec:no_tl_results}

\begin{figure*}
     \centering
     \begin{subfigure}[b]{0.45\textwidth}
         \centering
         \includegraphics[width=\textwidth]{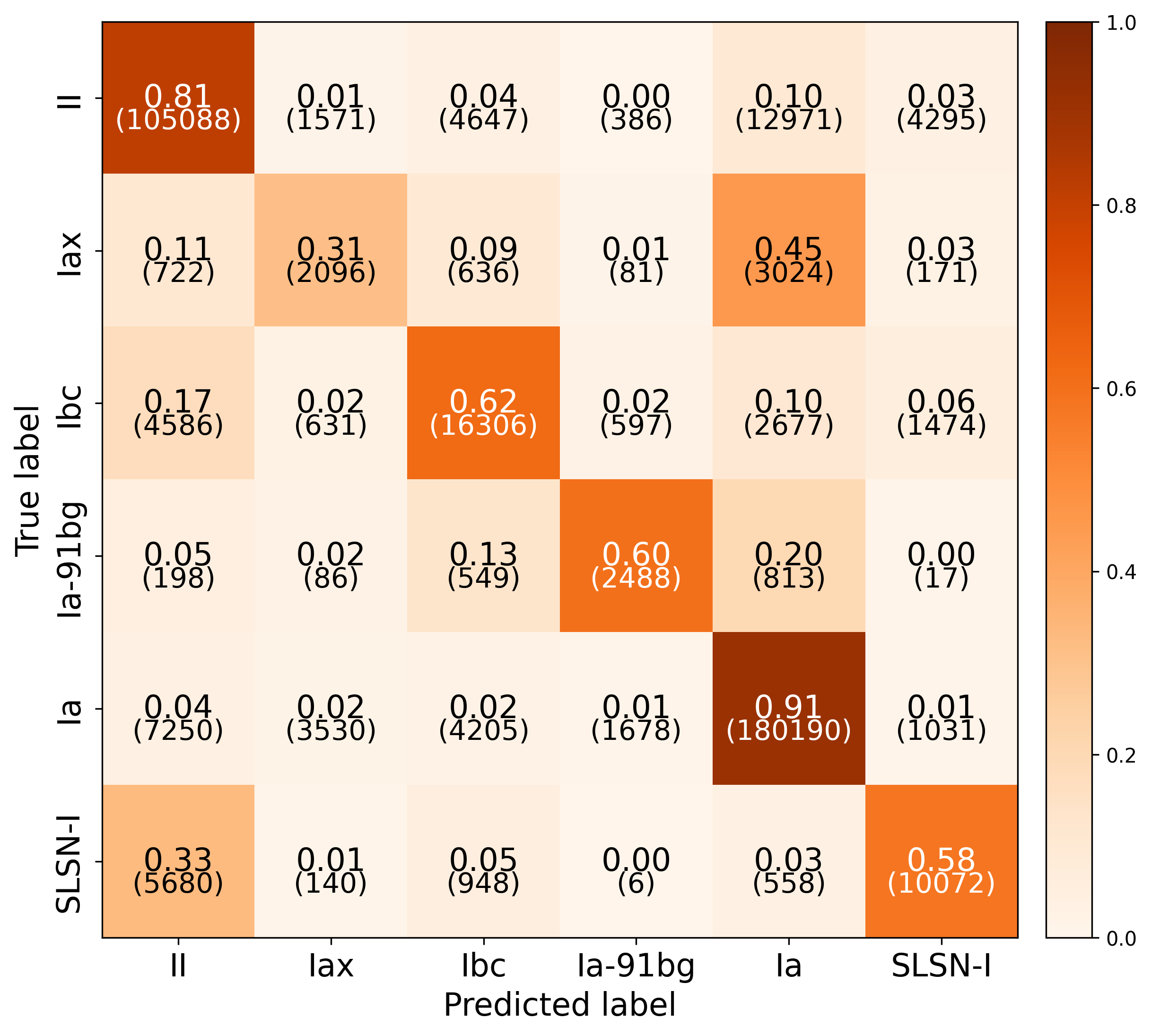}
         \caption{Original training set, no redshift}
         \label{fig:pl_o}
     \end{subfigure}
     \hfill
     \begin{subfigure}[b]{0.45\textwidth}
         \centering
         \includegraphics[width=\textwidth]{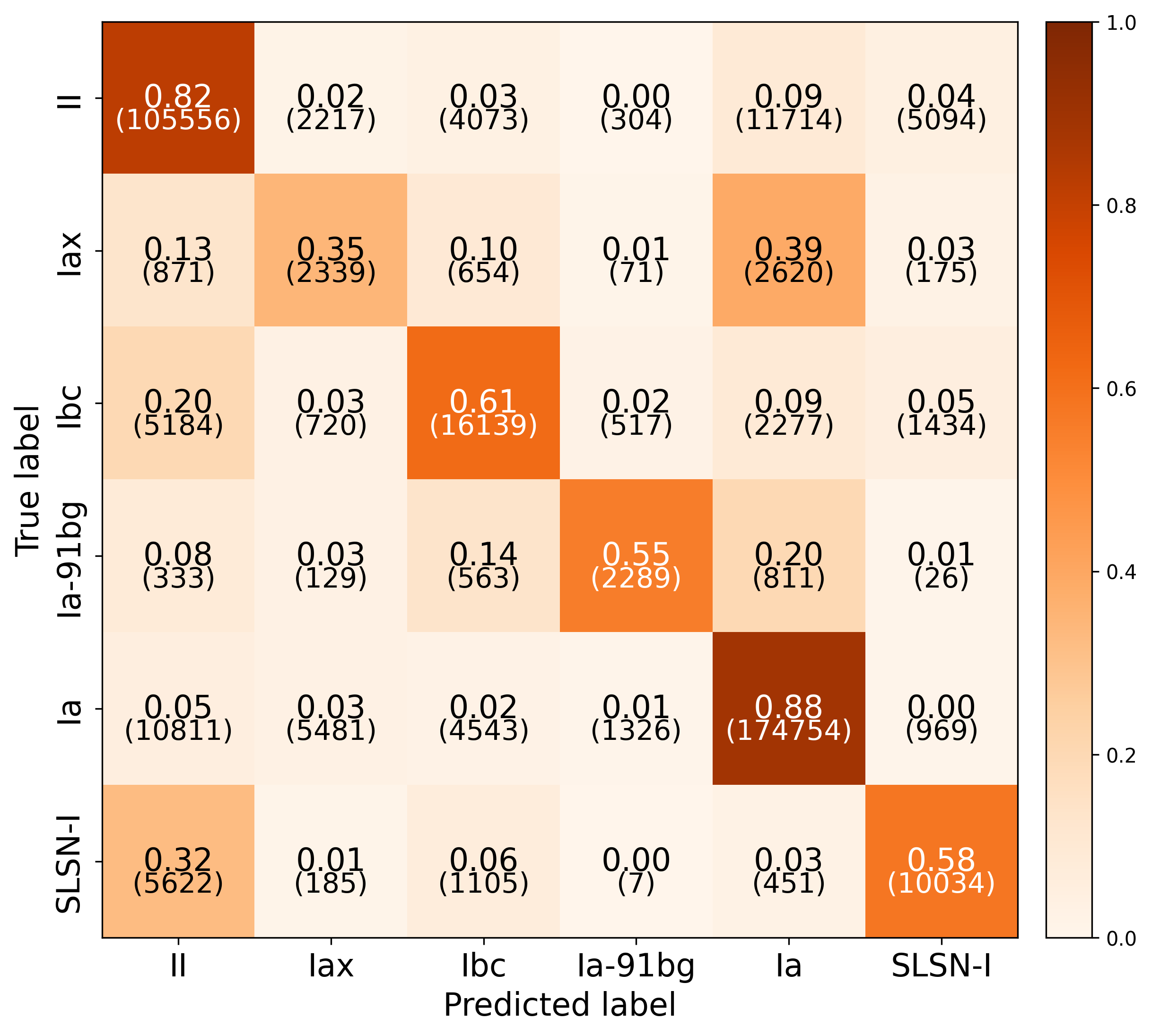}
         \caption{Original training set, with redshift}
         \label{fig:pl_oz}
     \end{subfigure}
    \newline
         \begin{subfigure}[b]{0.45\textwidth}
         \centering
         \includegraphics[width=\textwidth]{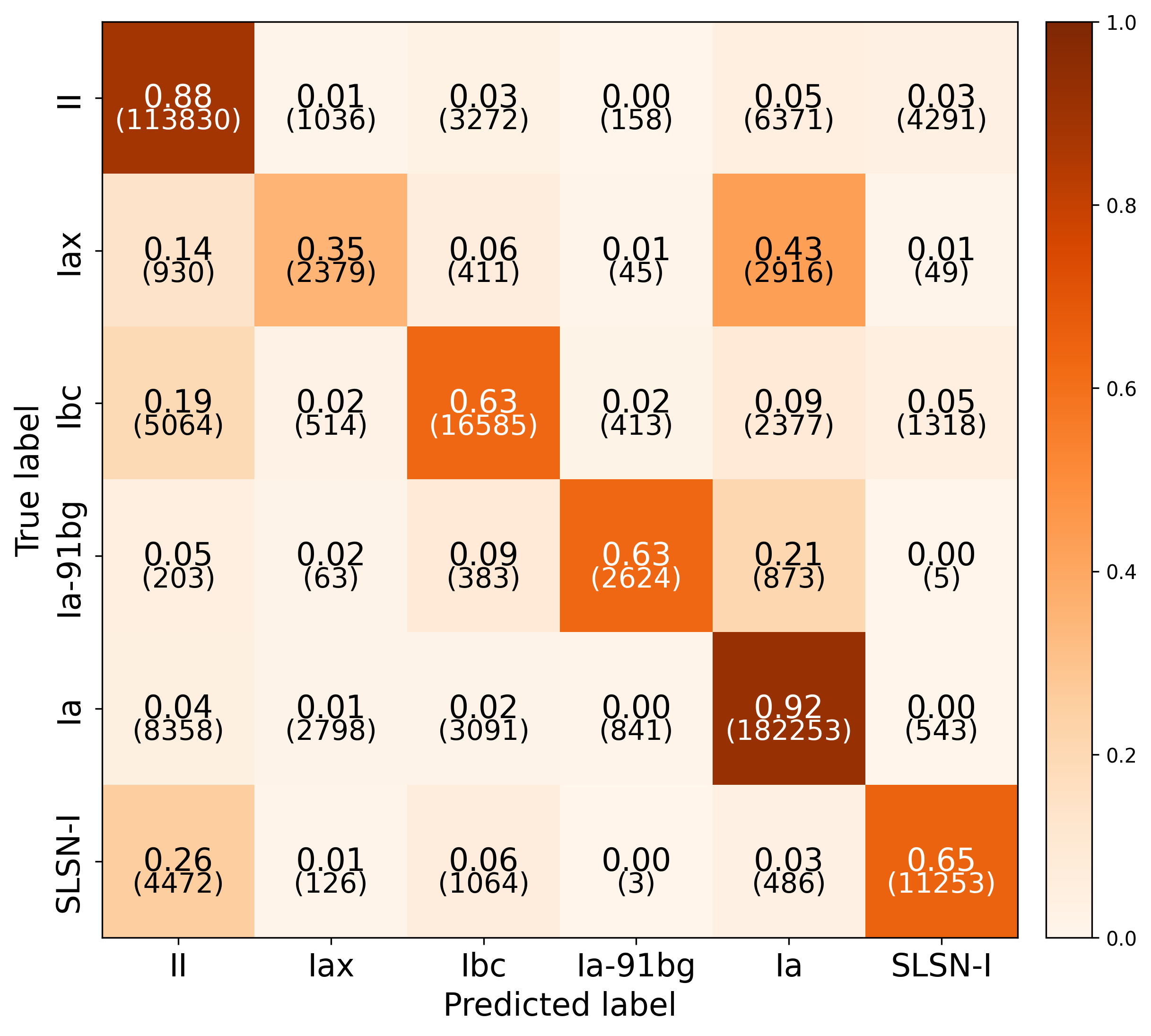}
         \caption{Augmented training set, no redshift}
         \label{fig:pl_a}
     \end{subfigure}
     \hfill
     \begin{subfigure}[b]{0.45\textwidth}
         \centering
         \includegraphics[width=\textwidth]{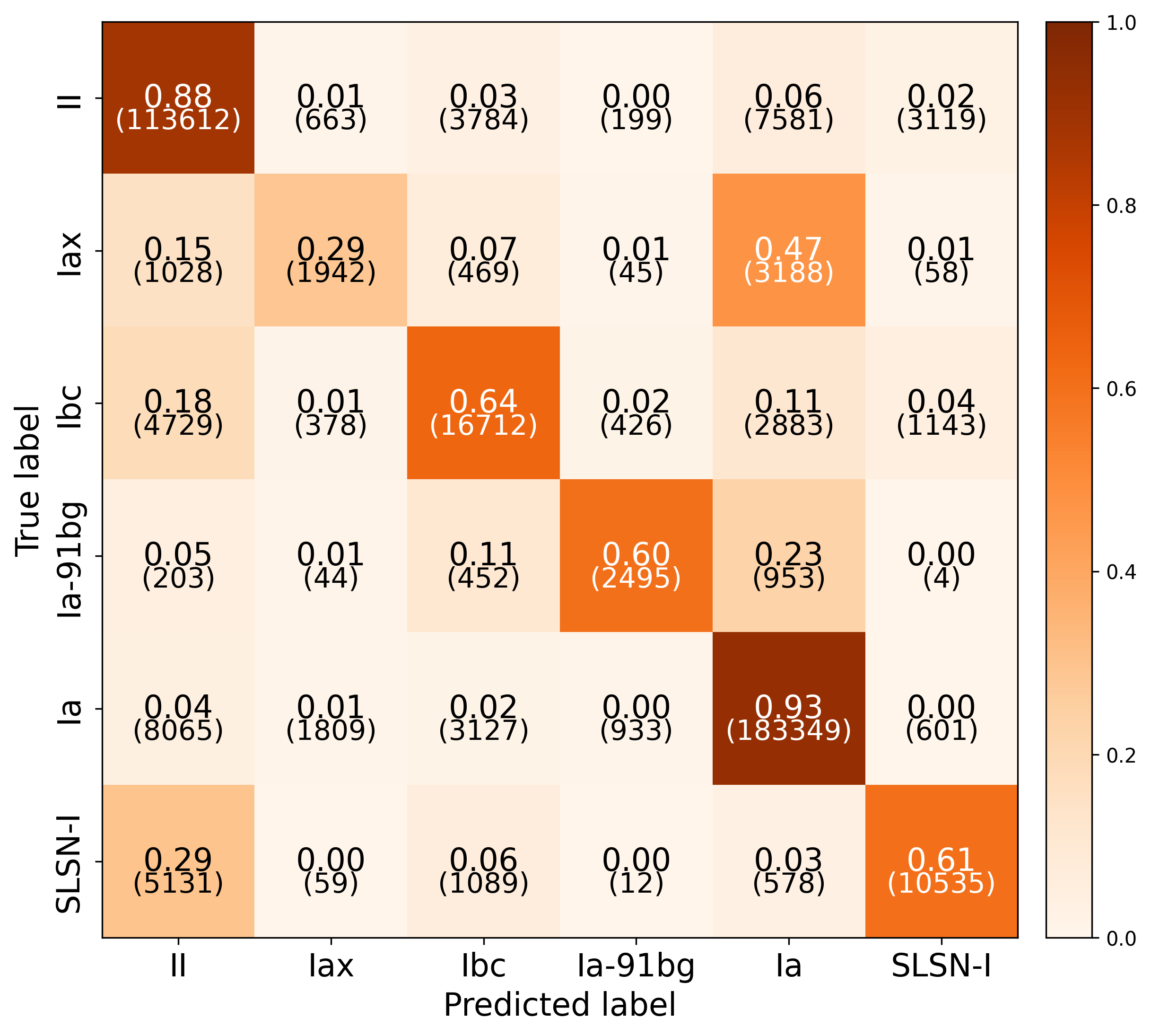}
         \caption{Augmented training set, with redshift}
         \label{fig:pl_az}
     \end{subfigure}
        \caption{Confusion matrices for models without transfer learning (trained only on the PLAsTiCC dataset), evaluated on the test set.}
        \label{fig:plasticc_confusion_matrices_notl}
\end{figure*}

After training, all models were evaluated on the test set. Figure \ref{fig:plasticc_confusion_matrices_notl} shows the confusion matrices for the models trained without transfer learning, using the original and augmented training set, with and without redshift information. Looking at the confusion matrices for the original training set, the model achieves good accuracy ($>80\%$) for type Ia and II supernovae, a medium level of accuracy for type Ibc, Ia-91bg, and SLSN-I ($>60\%$), and poor accuracy for type Iax supernovae. The biggest sources of confusion are type Iax and Ia-91bg being misclassified as Ia, and type Iax, Ibc, and SLSN-I being misclassified as type II. When redshift information is included, there is no major sign of improvement in performance.

When the augmented training set is used, there is a slight improvement in accuracy for type II supernovae (and increase of $\sim 7\%$ in accuracy), but no major change in performance in the other classes. Including redshift information does not have any significant improvement over the model without redshift information. There is an increase in the number of type Iax being misclassified as type Ia, and fewer SLSN-I being classified as type II.

\subsection{Models with transfer learning}
\label{subsec:with_tl_results}

\begin{figure*}
     \centering
     \begin{subfigure}[b]{0.45\textwidth}
         \centering
         \includegraphics[width=\textwidth]{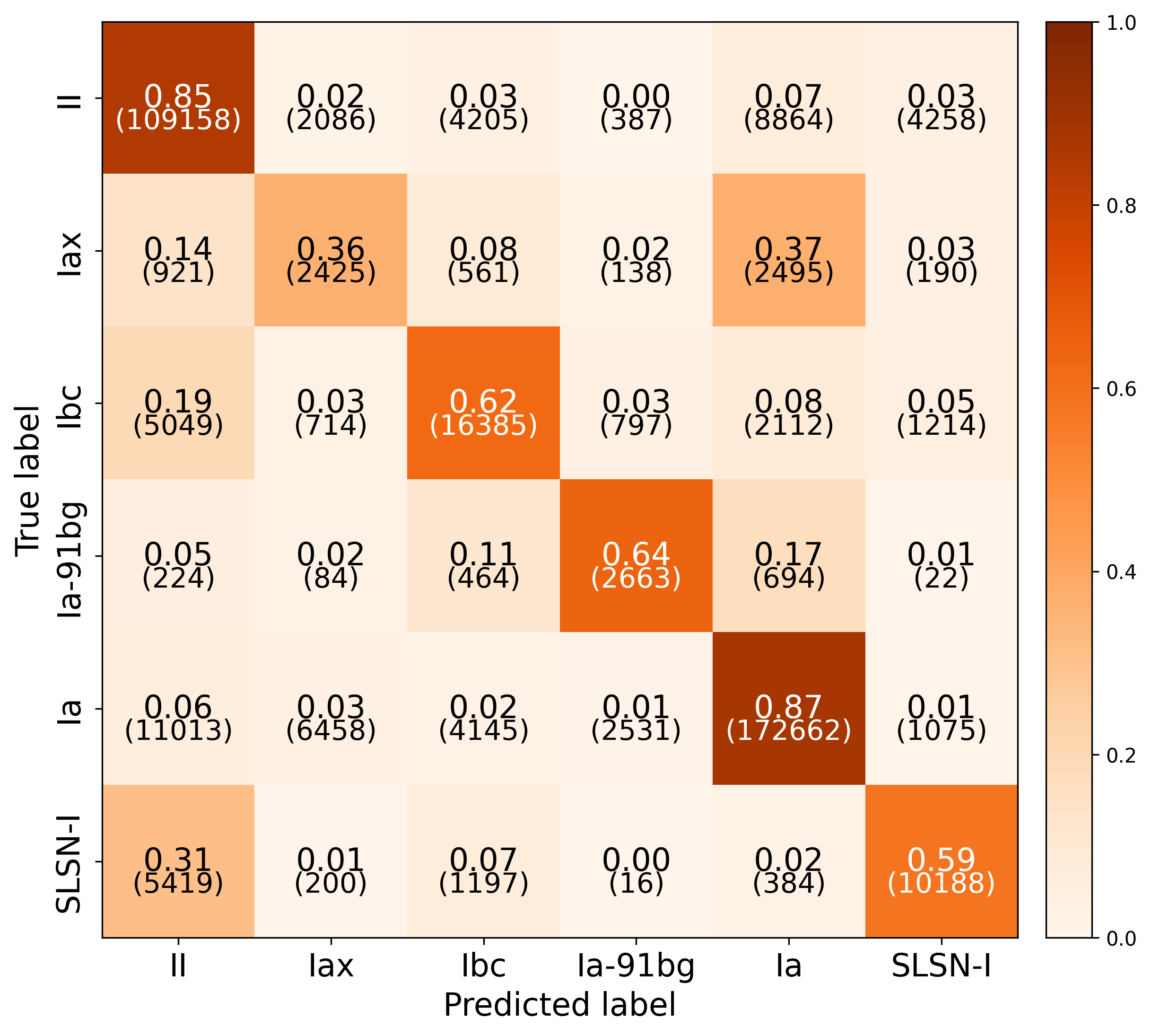}
         \caption{Original training set, no redshift}
         \label{fig:tl_o}
     \end{subfigure}
     \hfill
     \begin{subfigure}[b]{0.45\textwidth}
         \centering
         \includegraphics[width=\textwidth]{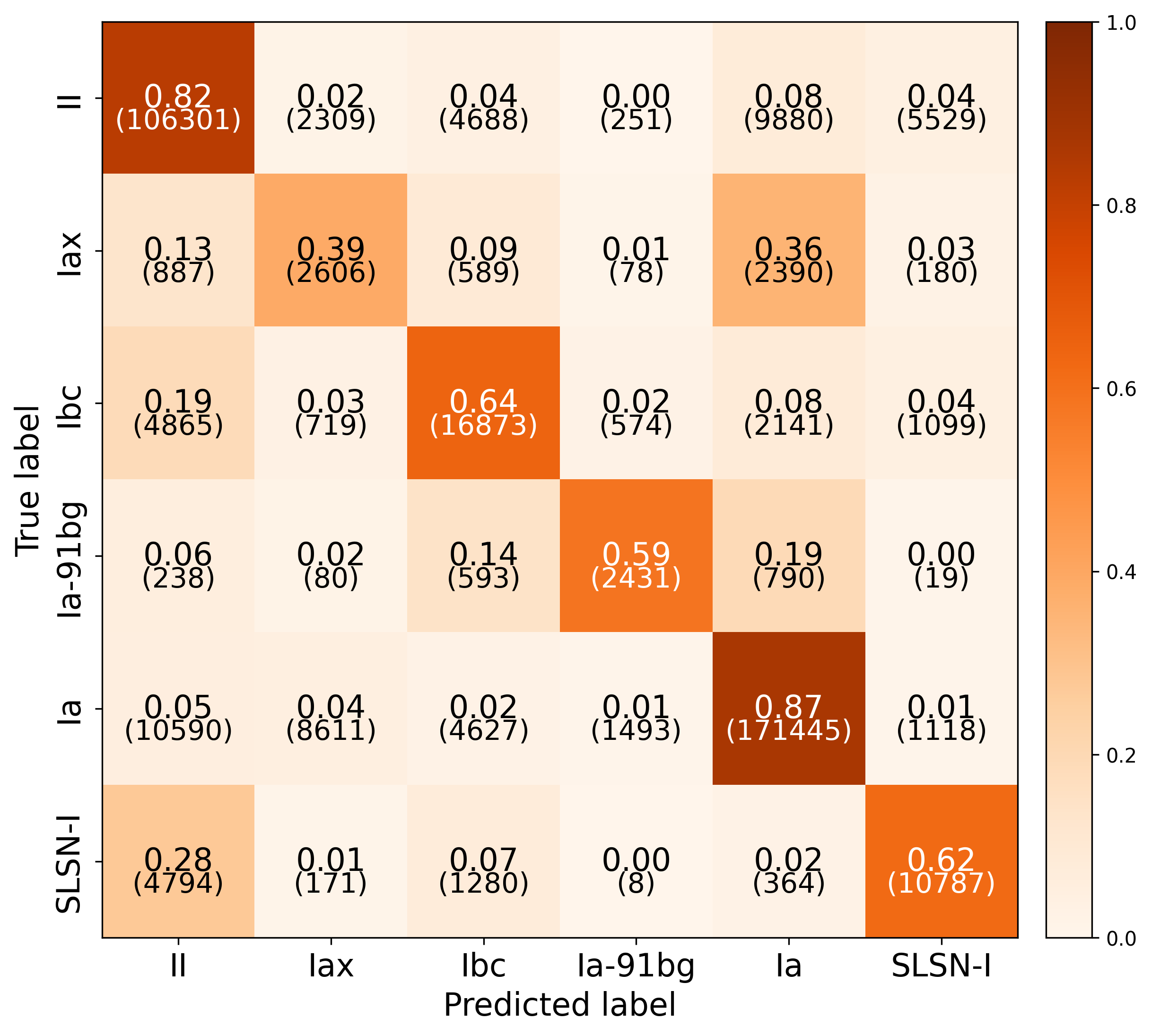}
         \caption{Original training set, with redshift}
         \label{fig:tl_oz}
     \end{subfigure}
    \newline
         \begin{subfigure}[b]{0.45\textwidth}
         \centering
         \includegraphics[width=\textwidth]{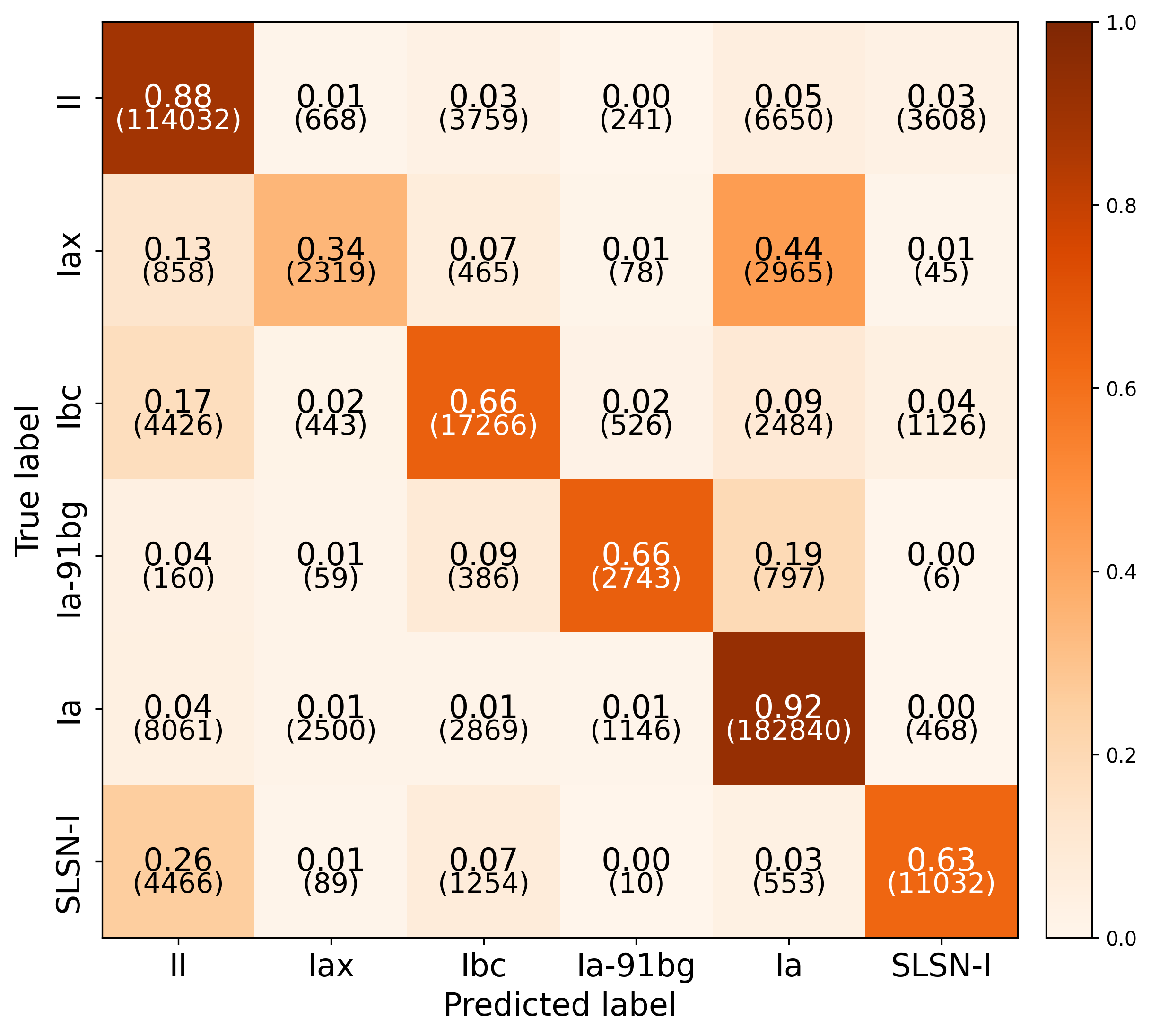}
         \caption{Augmented training set, no redshift}
         \label{fig:tl_a}
     \end{subfigure}
     \hfill
     \begin{subfigure}[b]{0.45\textwidth}
         \centering
         \includegraphics[width=\textwidth]{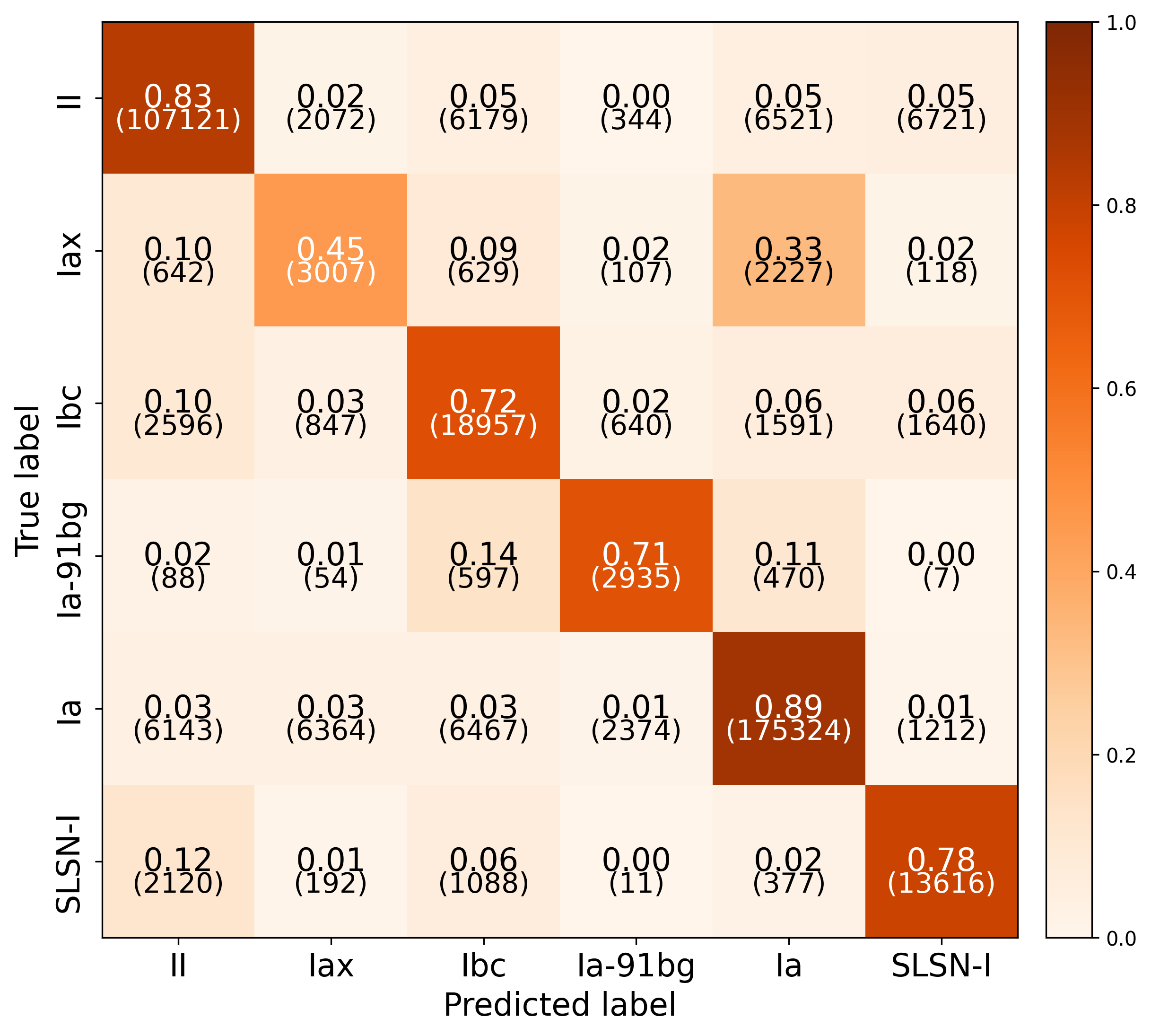}
         \caption{Augmented training set, with redshift}
         \label{fig:tl_az}
     \end{subfigure}
        \caption{Confusion matrices for models with transfer learning (using models trained on the Open Supernova Catalog dataset and then fine-tuned to the PLAsTiCC dataset), evaluated on the test set.}
        \label{fig:transfer_learning_confusion_matrices_withtl}
\end{figure*}

Figure \ref{fig:transfer_learning_confusion_matrices_withtl} shows the confusion matrices trained with transfer learning, using the original and augmented training set, with and without redshift information. For models trained on the original training set, both have slightly overall better performance over the same models without transfer learning, with an increase in accuracy by a few percent across most classes, and fewer misclassifications. When including redshift, there is no significant improvement in performance.

\begin{figure}
    \centering
    \includegraphics[width=0.45\textwidth]{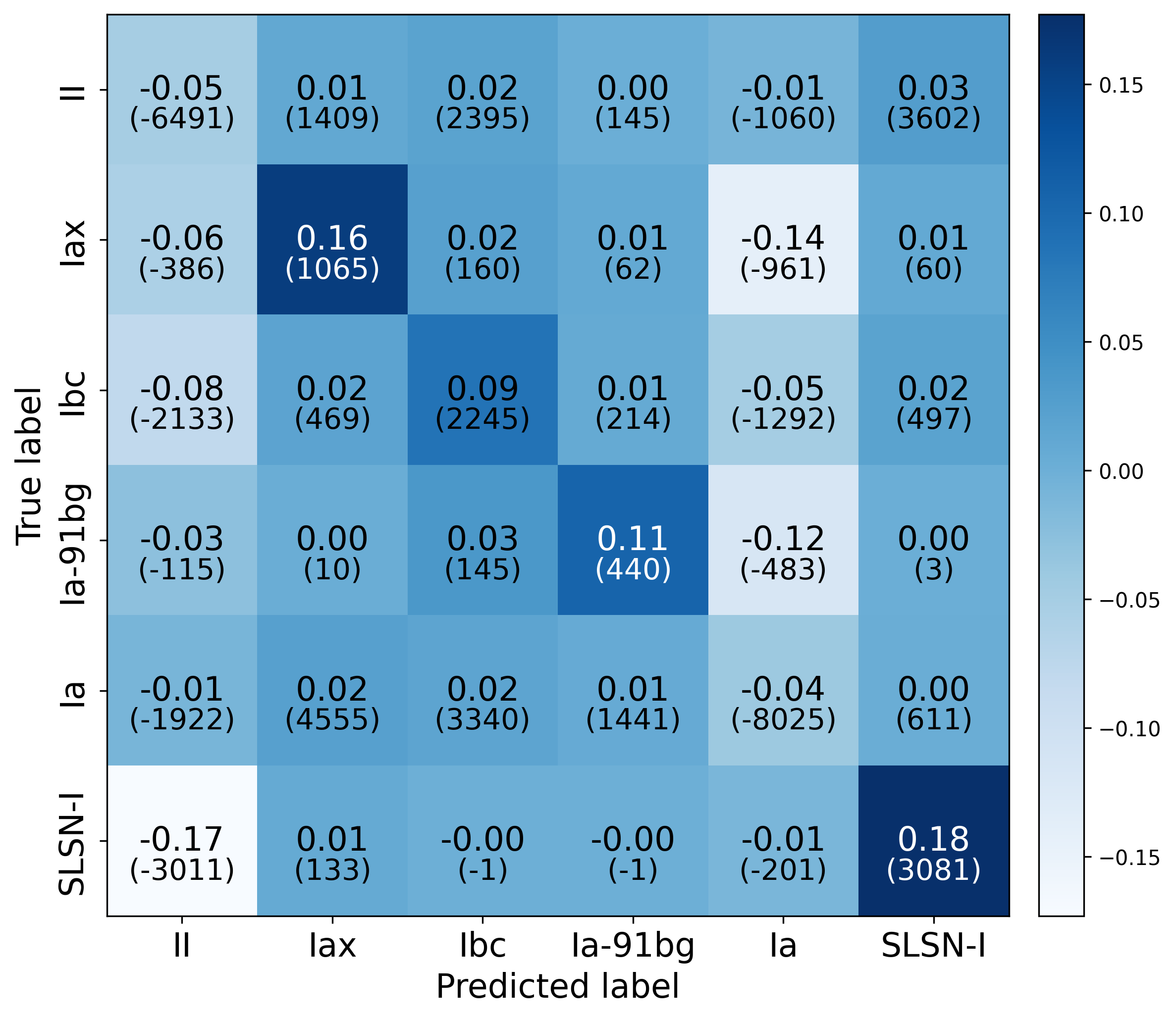}
    \caption{The difference between the confusion matrix for models trained on the augmented training set with redshift for with and without transfer learning. Positive values along the diagonal indicate an improvement when transfer learning is used. Negative values in the off diagonals indicate fewer misclassifications.}
    \label{fig:diff_cmtx}
\end{figure}

Looking at the models trained with the augmented training set, the performance for the model without redshift information is similar to the performance for the same model without transfer learning. When redshift information is included, the performance of the model with transfer learning is improved over the same model without transfer learning. There is good accuracy for type Ia and II supernovae ($>80\%$), and improved accuracy for type Ibc, Ia-91-bg, and SLSN-I ($>70\%$). There are fewer misclassifications overall ($<15\%$), and the model trained with transfer learning and redshift information achieves the best accuracy out of all models on type Iax ($45\%$). We plot the difference between the confusion matrices for the models trained with transfer learning and without, for the augmented training set with redshift in Figure \ref{fig:diff_cmtx}, to illustrate the change in performance between the two models.

\begin{figure*}
     \centering
         \begin{subfigure}[b]{0.45\textwidth}
         \centering
         \includegraphics[width=\textwidth]{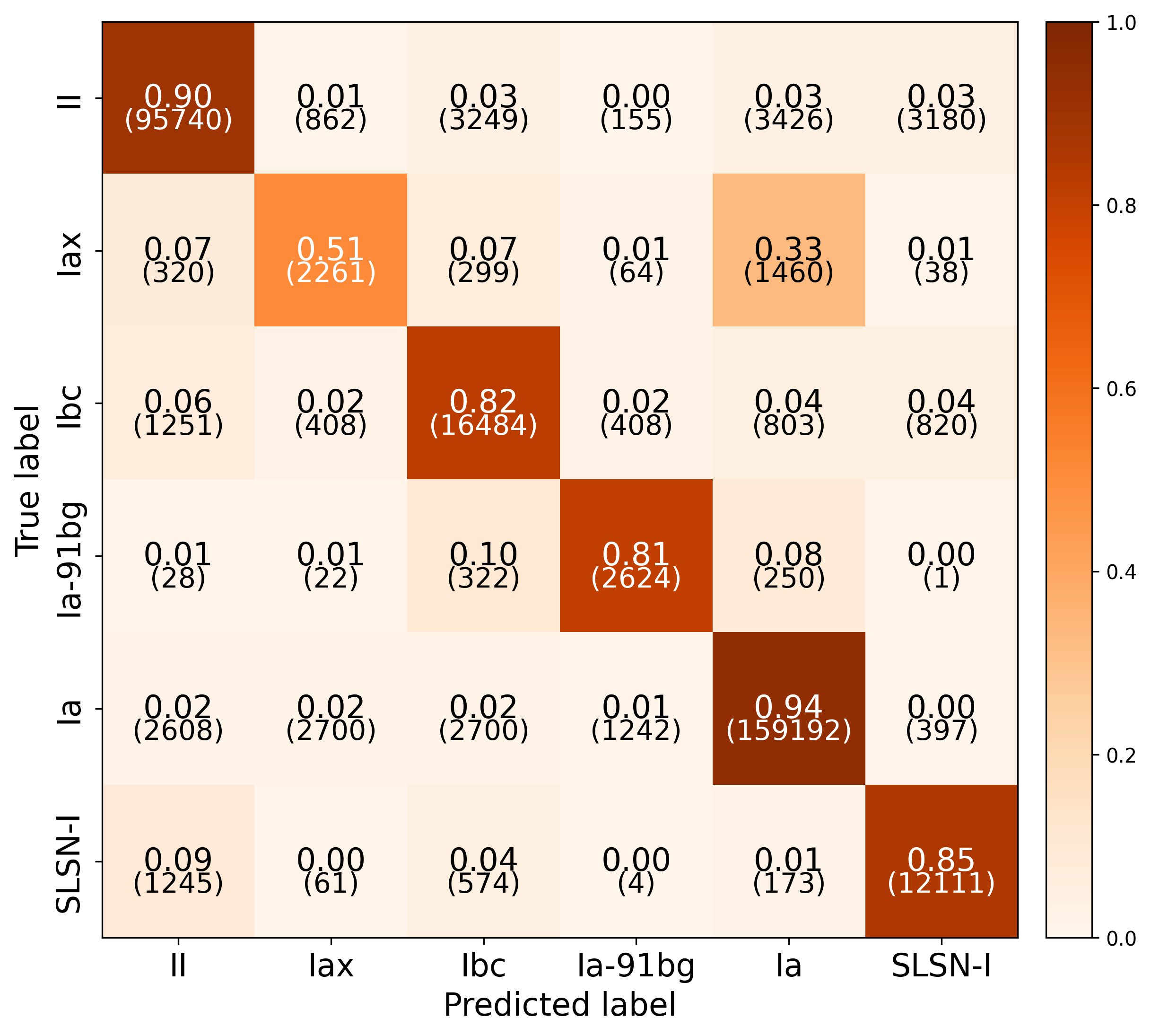}
         \caption{Augmented training set, with redshift, at a threshold of $0.7$}
         \label{fig:tl_az_07}
     \end{subfigure}
     \hfill
     \begin{subfigure}[b]{0.45\textwidth}
         \centering
         \includegraphics[width=\textwidth]{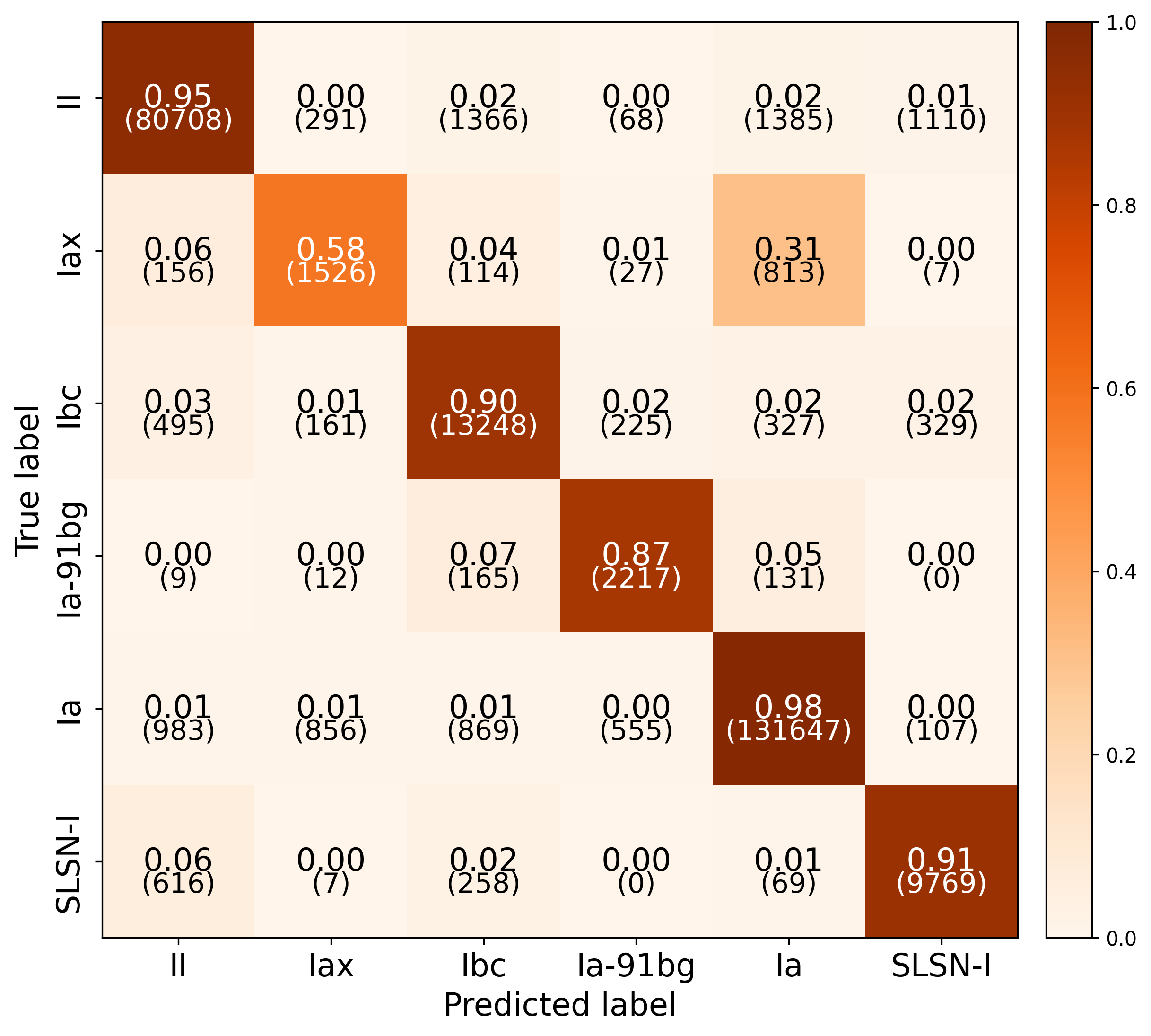}
         \caption{Augmented training set, with redshift, at a threshold of $0.9$}
         \label{fig:tl_az_09}
     \end{subfigure}
        \caption{Confusion matrices for the model with transfer learning, trained on the augmented training set with redshift at different thresholds.}
        \label{fig:threshold_confusion_matrices}
\end{figure*}

Table \ref{tab:summary_learning_table} shows the area under the receiver operating characteristic curve (AUC) score and the $F_1$ score for all trained models. The \cite{handtill2001} formulation is used to obtain the multi-class AUC scores presented in Table \ref{tab:summary_learning_table}. In both cases with models trained with and without transfer learning, including redshift yields an improvement in the AUC score, but not always an improvement in the $F_1$ score. A higher AUC score indicates that the model is able to produce fewer false positives, so when redshift is included the models are able to make predictions that have slightly less contamination at the small cost of not correctly classifying all true positive samples in each class.

We also examine how selecting a threshold for class membership reduces the number of false positives in each class. Since the model makes predictions by producing a list of scores that represent how likely an object belongs to a specific class, we can define a threshold score so that if the score is above the threshold then the object belongs to that class, and if it is below then it is considered to not belong to that class. Three threshold values are selected: 0.5, 0.7, and 0.9. We consider the model trained with transfer learning using the augmented training set and redshift information. For each threshold value, any predictions that are below the threshold are excluded (looking at the highest score out of the six classes). Table \ref{tab:transfer_learning_threshold_table} lists the AUC score, $F_1$ score, and the fraction of samples retained at the different threshold values. Figure \ref{fig:threshold_confusion_matrices} shows the confusion matrices for thresholds at 0.7 and 0.9. As the threshold value increases, the AUC score and $F_1$ score improves but the fraction of samples retained decreases. 

\begin{table*}
    \centering
    \begin{tabular}{c|ccc}
        \toprule
         &\textbf{Model} & \textbf{AUC} & {$\pmb{\mathrm{F}_{1}}$} \\
         \hline
         \multirow{4}{*}{No transfer learning}&Original, no redshift & 0.893 & 0.624 \\
         &Original, with redshift & 0.895 & 0.614 \\
         &Augmented, no redshift & 0.924 & 0.680 \\
         &Augmented, with redshift & 0.921 & 0.669 \\
         \hline
         \multirow{4}{*}{With transfer learning}&Original, no redshift & 0.901 & 0.617 \\
         &Original, with redshift & 0.915 & 0.620 \\
         &Augmented, no redshift & 0.925 & 0.683 \\
         &Augmented, with redshift & 0.945 & 0.657 \\
         \bottomrule
    \end{tabular}
    \caption{AUC and $\mathrm{F}_1$ scores, trained on the original and augmented training sets with and without redshift, for both with and without transfer learning.}
    \label{tab:summary_learning_table}
\end{table*}

\begin{table}
    \centering
    \begin{tabular}{c|ccc}
        \toprule
         \textbf{Threshold} & \textbf{AUC} & {$\pmb{\mathrm{F}_{1}}$} & \textbf{Fraction retained}\\
         \hline
         0.5 & 0.950 & 0.684 & 95.6\% \\
         0.7 & 0.960 & 0.752 & 83.2\% \\
         0.9 & 0.971 & 0.838 & 65.7\% \\
         \bottomrule
    \end{tabular}
    \caption{AUC and $\mathrm{F}_1$ scores for the transfer learning model trained on the augmented training set with redshift, evaluated at different probability thresholds. The column on the right shows the fraction of the test set retained when discarding predictions that are below the threshold.}
    \label{tab:transfer_learning_threshold_table}
\end{table}

\subsection{Limited training set performance}
\label{subsec:partial_trainset}

To further investigate the impact transfer learning would have at the start-up phase of a new survey (i.e. when there is a very small sample of labelled data), we compare the classification performance of models with and without transfer learning when trained on 10\%, 25\%, and 50\% of the original PLAsTiCC training set. Table \ref{tab:partial_trainset_table} shows the AUC and $F_1$ scores for these models, when trained with and without redshift information.

\begin{table}
    \centering
    \begin{tabular}{c|ccc}
        \toprule
         &\textbf{Model} & \textbf{AUC} & {$\pmb{\mathrm{F}_{1}}$} \\
         \hline
         \multirow{6}{*}{With transfer learning}&50\% of original, no redshift & 0.888 & 0.582 \\
         &50\% of original, with redshift & 0.899 & 0.577 \\
         &25\% of original, no redshift & 0.869& 0.574 \\
         &25\% of original, with redshift & 0.884 & 0.573 \\
         &10\% of original, no redshift & 0.827& 0.520 \\
         &10\% of original, with redshift & 0.844 & 0.510 \\
         \hline
         \multirow{6}{*}{No transfer learning}&50\% of original, no redshift & 0.880 & 0.604 \\
         &50\% of original, with redshift & 0.892 & 0.600 \\
         &25\% of original, no redshift & 0.880& 0.579 \\
         &25\% of original, with redshift & 0.890 & 0.577 \\
         &10\% of original, no redshift & 0.836& 0.513 \\
         &10\% of original, with redshift & 0.832 & 0.523 \\
         \bottomrule
    \end{tabular}
    \caption{AUC and $\mathrm{F}_1$ scores, for models trained on different fractions of the original PLAsTiCC training set with and without transfer learning.}
    \label{tab:partial_trainset_table}
\end{table}

From Table \ref{tab:partial_trainset_table}, we can see that the AUC and $F_1$ scores improve with an increased training set size for both with and without transfer learning. Overall, including redshift information improves performance across all models. An interesting point is that when no transfer learning is used, classification performance remains comparable to when transfer learning is used, and models trained on 50\% of the original training set shown an increase in $F_1$ score with transfer learning compared to when no transfer learning is used.

\begin{figure}
    \centering
    \includegraphics[width=0.48\textwidth]{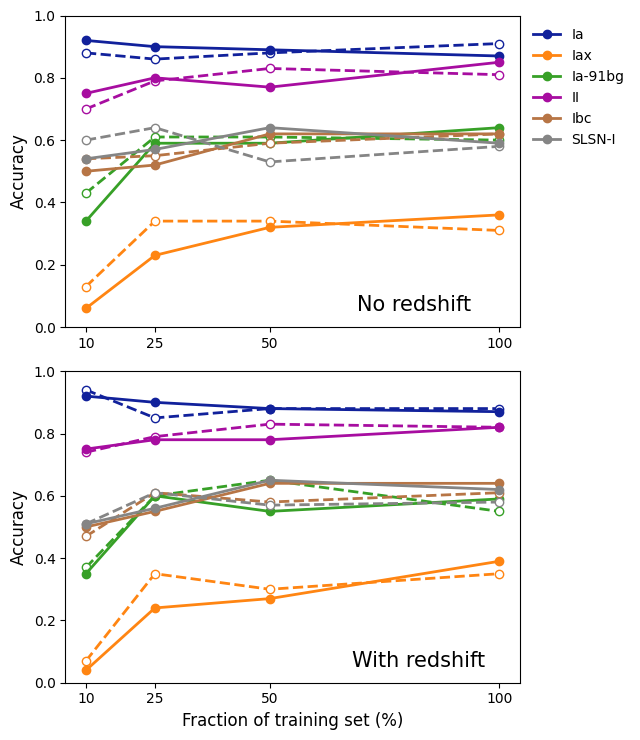}
    \caption{PLAsTiCC test accuracies across the six classes for different fractions of the original PLAsTiCC training set. The solid lines indicate accuracies for models with transfer learning and the dashed lines indicate accuracies for models without transfer learning.}
    \label{fig:plasticc_partial_accuracy}
\end{figure}

To see how transfer learning impacts classification for individual supernova classes, we plot how the accuracy varies for each class as a function on training set size with and without redshift information in Figure \ref{fig:plasticc_partial_accuracy}. All models achieve the best accuracies for type Ia and Ibc supernovae, and the worst accuracy for type Iax. There is no significant improvement in performance across all classes when transfer learning is used compared to training on the original PLAsTiCC training set alone. What is notable is that for type Iax and Ia-91bg the classification accuracy is worse with transfer learning at smaller training set sizes. This could be due to the fact that the Open Supernova Catalog data contains many type Ia examples, and there is a small number of type Iax and 1a-91bg examples in the PLAsTiCC training set. We also see that as the PLAsTiCC training set grows, the accuracy for type Ia degrades while the accuracies for all other classes improve. This may arise from the fact that as the number of examples from non-Ia classes increases, the model learns to better classify non-Ia supernovae at the cost of misclassifying some type Ia supernovae.

\section{Discussion \& Conclusions}
\label{sec:discussion}


\subsection{Classifying supernovae from multiple surveys}
\label{subsec:classifying_sne}


In order to classify the heterogeneous supernova light curve dataset, we use a two-dimensional Gaussian process to model the light curves, and create a flux heatmap image for each light curve where each pixel in the image represents flux, as a function of time and wavelength. We also show that in the case where a supernova has good photometric coverage in multiple filters (measuring the flux at different wavelengths), the Gaussian process can be used to generate a low-resolution spectra of the supernova. Comparing the real spectra and Gaussian process generated spectra of supernova iPTF13bvn, we find that the two are comparable and the generated spectra does resemble the real spectra. It is not guaranteed, however, that all supernovae will have the same quality of photometric observations. Out of the original $\sim80,000$ supernova light curves from the Open Supernova Catalog, only 6,330 were used to generate flux heatmaps with a a two-dimensional Gaussian process after selection cuts. A larger sample could be used, but at the cost of lower quality light curves (i.e. poor sampling in time and lack of multi-colour observations) which may result in poor fitting with Gaussian processes.

We used a convolutional neural network to classify the supernova flux heatmaps, since the data is in a grid format which is well suited for the convolution operations carried out in the neural network. The model is able to classify type Ia and II supernovae with good accuracy, but the class imbalance in the dataset presents a challenge for classifying type Ibc supernovae, since it is the class with the smallest number of samples and is not well represented in the training set. Deep learning approaches benefit from having a large dataset to learn from, and we note that the Open Supernova Catalog dataset is rather small for a deep learning application with less than 4,000 samples in the training set. A future work may benefit from using a larger training set using flux heatmaps generated from simulated supernova light curves. It may also be interesting to investigate how the number of filters in a light curve (i.e. wavelength coverage) affects the flux heatmap.

\subsection{Transfer learning for future surveys}
\label{subsec:transfer_learning_future}

We used a subset of $397,990$ supernova light curves from the PLAsTiCC dataset \citep{PLAsTiCC2018}, and use the two-dimensional Gaussian process to generate flux heatmaps from the light curves. For the PLAsTiCC dataset we split the data into six classes, presenting a different classification task than the one for the Open Supernova Catalog dataset. The original training set (containing 2,398 SNe) and an augmented training set (containing 14,914) are used. Typically in most machine learning and deep learning methods, the training set is larger than the test set. Here, we use training sets that are much smaller than the test set to emulate the case where there is a scarcity of labelled data (the training set) and a large amount of unlabelled data (the test set).

In Section \ref{sec:plasticc_results}, we demonstrate that it is possible to transfer knowledge between two different domains (Open Supernova Catalog data and PLAsTiCC data) and two different classification tasks (three classes to six classes). The use of transfer learning shows a small improvement over when no transfer learning is used (and the model is only trained on the PLAsTiCC training set). We find the best increase in performance comes when redshift information is included and the augmented training set is used. It is possible to obtain better classifications with fewer misclassifications between classes when a threshold is used to remove `unconfident' classifications provided by the classifier. Looking at the impact transfer learning has when there is a small labelled training set, we find that there is a slight improvement for well-sampled classes (e.g. type Ia and II supernovae), but provides no benefit for classes that have very few examples. 

A limitation of the two-dimensional Gaussian process approach used in this work is the requirement for sufficiently good coverage across multiple photometric bands and also in time. For future surveys such as LSST, this is dependent on the choice of observing strategy to provide a good enough cadence and wavelength coverage. The use of a two-dimensional Gaussian process also relies on the full supernova light curve to create a good flux heatmap representation, which is still useful for the retrospective classification of supernovae to create samples for population studies. 


In this paper, we present an approach to classify Open Supernova Catalog light curves from multiple surveys with a convolutional neural network by using a two-dimensional Gaussian process to generate an image representation of supernova light curves. We find that using this method achieves good classification when there is good representation of the data in the training set. In the case of type Ibc classification, the performance is poor since there is a lack of representation of type Ibc supernovae in the training set. For classification tasks, it is important to have a good representative training set with good coverage in feature space for all classes so that a model is able to learn the feature-class relationship to make robust classifications.

We then investigate the usefulness of transfer learning in the context of future surveys where there may be a lack of labelled data to form a training set with which to train classifiers. The use of transfer learning shows a small improvement in classifiers compared to when no transfer learning is used when classifying PLAsTiCC supernova light curves. The addition of contextual information such as redshift and an augmented training set provided the best improvement in classification performance, highlighting the importance of a representative training set and the benefits of incorporating contextual information when classifying light curves. In the case of using transfer learning when there is a very small labelled training set, it may be useful to adapt a model that has been trained on a representative training set to account for class imbalance.

The methods presented in this paper could also be extended to classifying light curves of other non-supernova objects (such as variable stars, flare events, and AGN). The flux heatmaps generated with the two-dimensional Gaussian process could be used with a different neural network architecture such as a recurrent neural network, where each the input at each time step is a single column of the heatmap representing the flux interpolated along wavelength. This would allow classifications to be obtained with time, and also be used to classify partial light curves (unlike the full light curves used in this paper), where the Gaussian process is used to interpolate the light curves up to the most recent observation as in \cite{qu2021earlytime}. 

A classification model that is agnostic to the different filters used across different surveys would be useful in the near future of time-domain astronomy. New objects observed by surveys such as LSST with the Vera Rubin Observatory could trigger follow-up observations by various instruments world wide, which could be ingested by such a model to provide fast early-time classifications to identify good candidates for time-sensitive observations. 

\section*{Acknowledgements}

The research of U.F.B and J.R.M are funded through a Royal Society PhD studentship (Royal Society Enhancement Award RGF\textbackslash EA\textbackslash180234) and STFC grant ST/V000853/1 respectively.


\section*{Data Availability}

The work presented in this paper makes use of publicly available data from the Open Supernova Catalog (https://github.com/astrocatalogs) and the unblinded PLAsTiCC Classification Challenge dataset (http://doi.org/10.5281/zenodo.2539456).
 



\bibliographystyle{mnras}
\bibliography{references}




\appendix

\section{Magnitude conversions}

Conversions between Swift magnitudes, Vega magnitudes, and the Carnegie Supernova Project magnitude system into AB magnitudes are showin in Tables \ref{tab:swift_AB}, \ref{tab:CSP_AB}, and \ref{tab:CSP_AB}. 

\begin{table}
    \centering
    \begin{tabular}{lc}
    \toprule
    \textbf{Filter} & \textbf{AB - Vega}\\
    \hline
    $V$ &	-0.01\\
    $B$ &	-0.13\\
    $U$ &	+1.02\\
    UVW1 & +1.51\\
    UVM2 &	+1.69\\
    UVW2 &	+1.73\\
    \bottomrule
    \end{tabular}
    \caption{Conversion table for Swift magnitudes given in the Vega system into AB magnitudes, obtained from \citet{breeveld2011}}
    \label{tab:swift_AB}
\end{table}

\begin{table}
    \centering
    \begin{tabular}{lc}
    \toprule
    \textbf{Filter} & \textbf{AB - Vega}\\
    \hline
    $U$ &	0.79\\
    $B$ &	-0.09\\
    $V$ &	0.02\\
    $R$ &	0.21\\
    $I$ &	0.45\\
    $u$ &	0.91\\
    $g$ &	-0.08\\
    $r$ &	0.16\\
    $i$ &	0.37\\
    $z$ &	0.54\\
    $J$ &	0.91\\
    $H$ &	1.39\\
    $K_{\mathrm{s}}$ & 1.85\\
    \bottomrule
    \end{tabular}
    \caption{Conversion table for Vega magnitudes into AB magnitudes, obtained from \citet{blanton2007}}
    \label{tab:vega_AB}
\end{table}

\begin{table}
    \centering
    \begin{tabular}{lc}
    \toprule
    \textbf{Filter} & \textbf{AB - CSP}\\
    \hline
    $u$ &	-0.06\\
    $g$ &	-0.02\\
    $r$ &	-0.01\\
    $i$ &	0.00\\
    $B$ &	-0.013\\
    $V$ &	-0.02\\
    $Y_\mathrm{RC}$ &	0.63\\
    $J$ &	0.90\\
    $H_\mathrm{RC}$ &	1.34\\
    $Y_\mathrm{WIRC}$ &	0.64\\
    $J_\mathrm{WIRC}$ &	0.90\\
    $H_\mathrm{WIRC}$ &	1.34\\
    \bottomrule
    \end{tabular}
    \caption{Conversion table for magnitudes given in the Carnegie Supernova Project (CSP) system into AB magnitudes, obtained from \citet{krisciunas2017}}
    \label{tab:CSP_AB}
\end{table}

\section{Transfer learning training}
\label{app:transfer_learning}

\begin{figure*}
     \centering
     \begin{subfigure}[b]{0.45\textwidth}
         \centering
         \includegraphics[width=\textwidth]{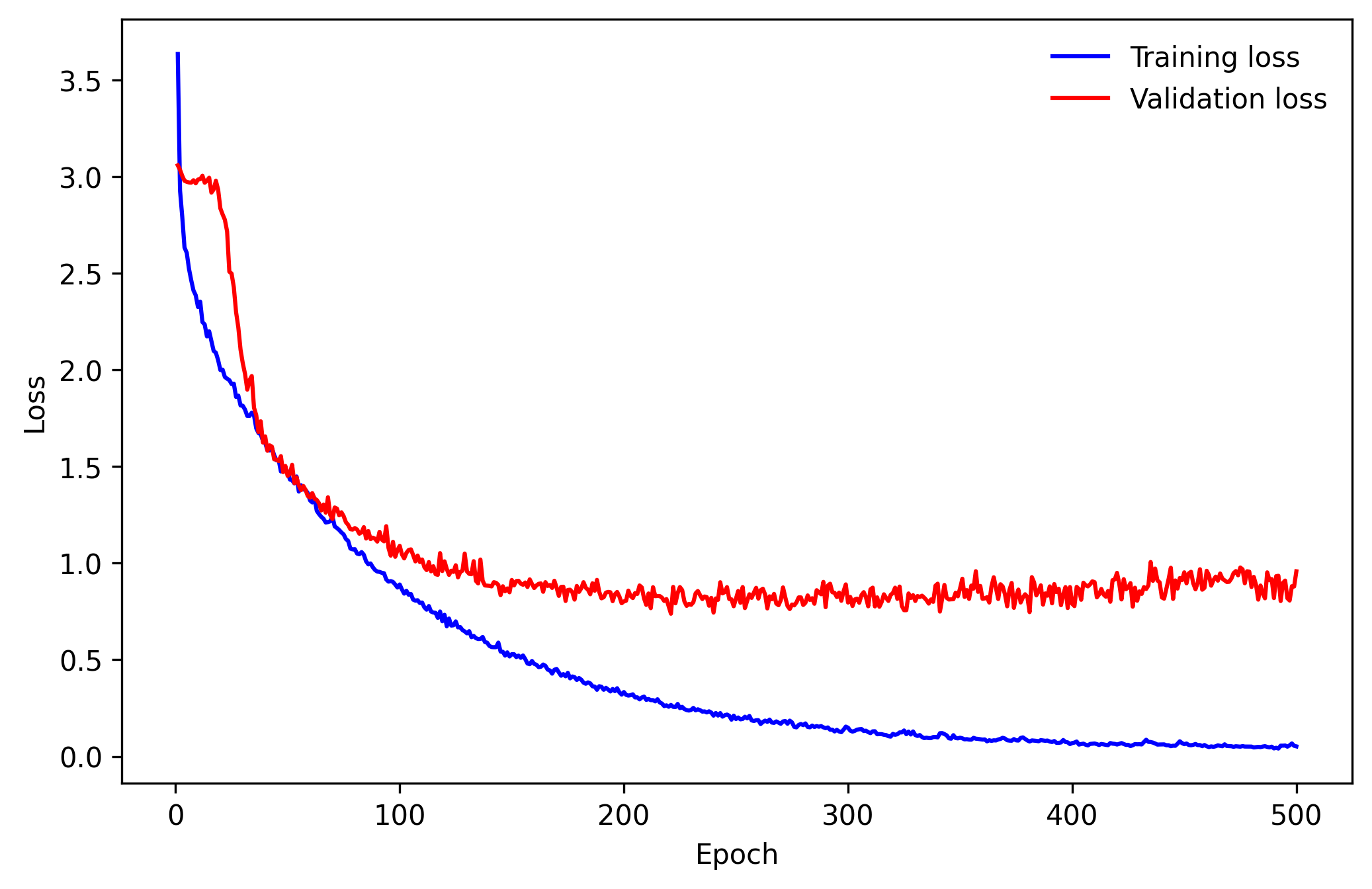}
         \caption{Original training set, no redshift}
         \label{fig:pl_o_hist_ntl}
     \end{subfigure}
     \hfill
     \begin{subfigure}[b]{0.45\textwidth}
         \centering
         \includegraphics[width=\textwidth]{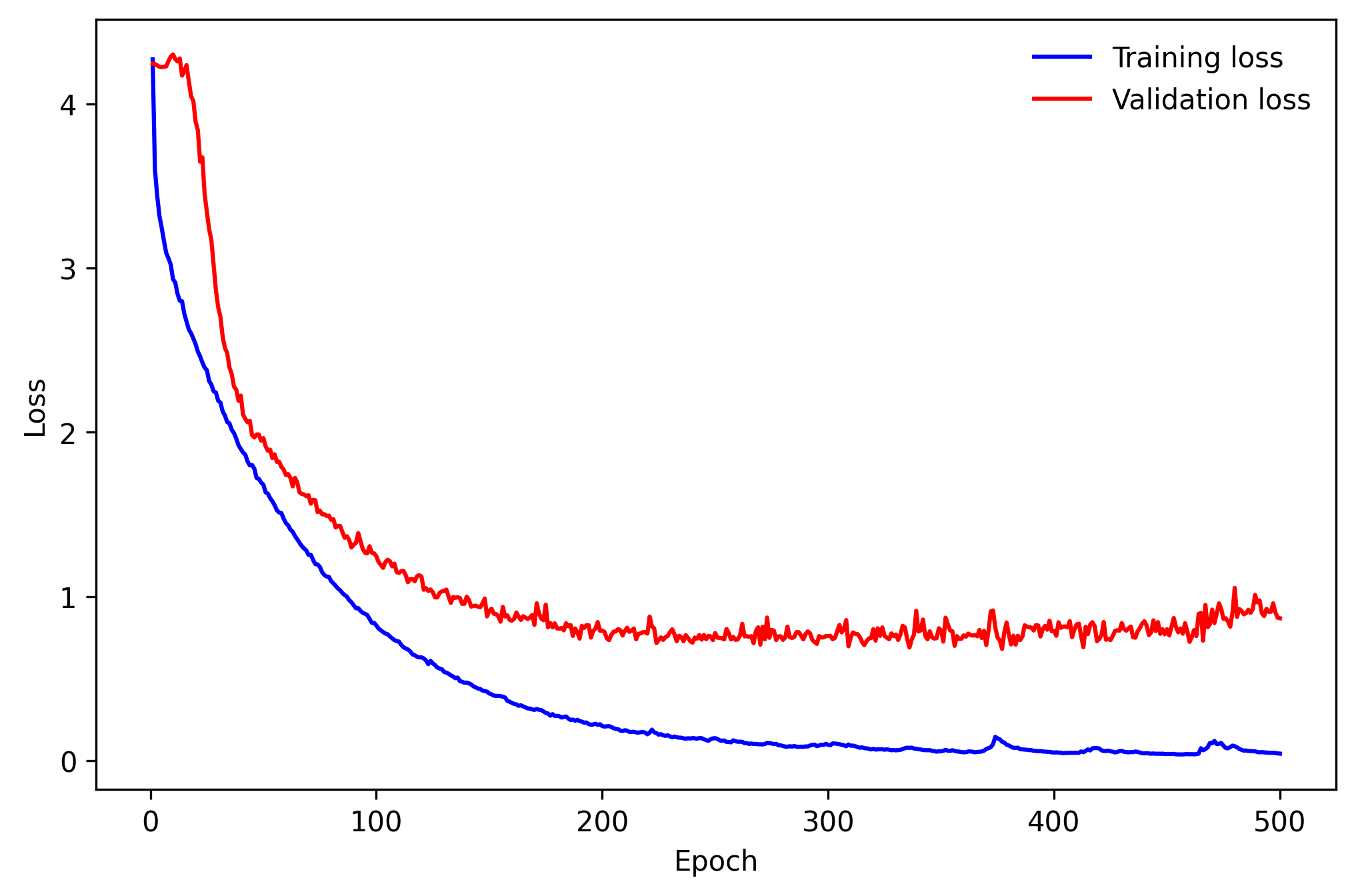}
         \caption{Original training set, with redshift}
         \label{fig:pl_oz_hist_ntl}
     \end{subfigure}
    \newline
         \begin{subfigure}[b]{0.45\textwidth}
         \centering
         \includegraphics[width=\textwidth]{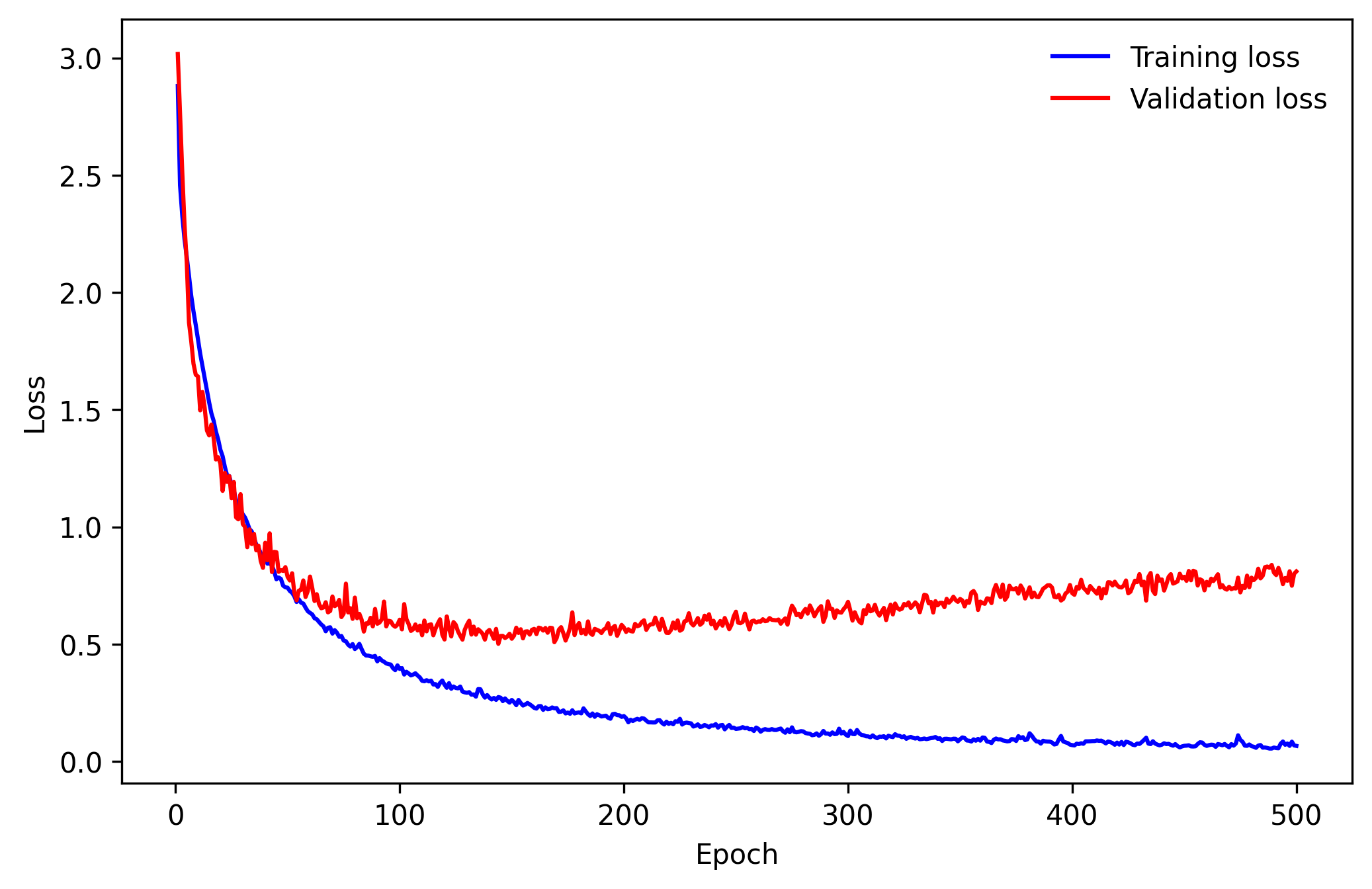}
         \caption{Augmented training set, no redshift}
         \label{fig:pl_a_hist_ntl}
     \end{subfigure}
     \hfill
     \begin{subfigure}[b]{0.45\textwidth}
         \centering
         \includegraphics[width=\textwidth]{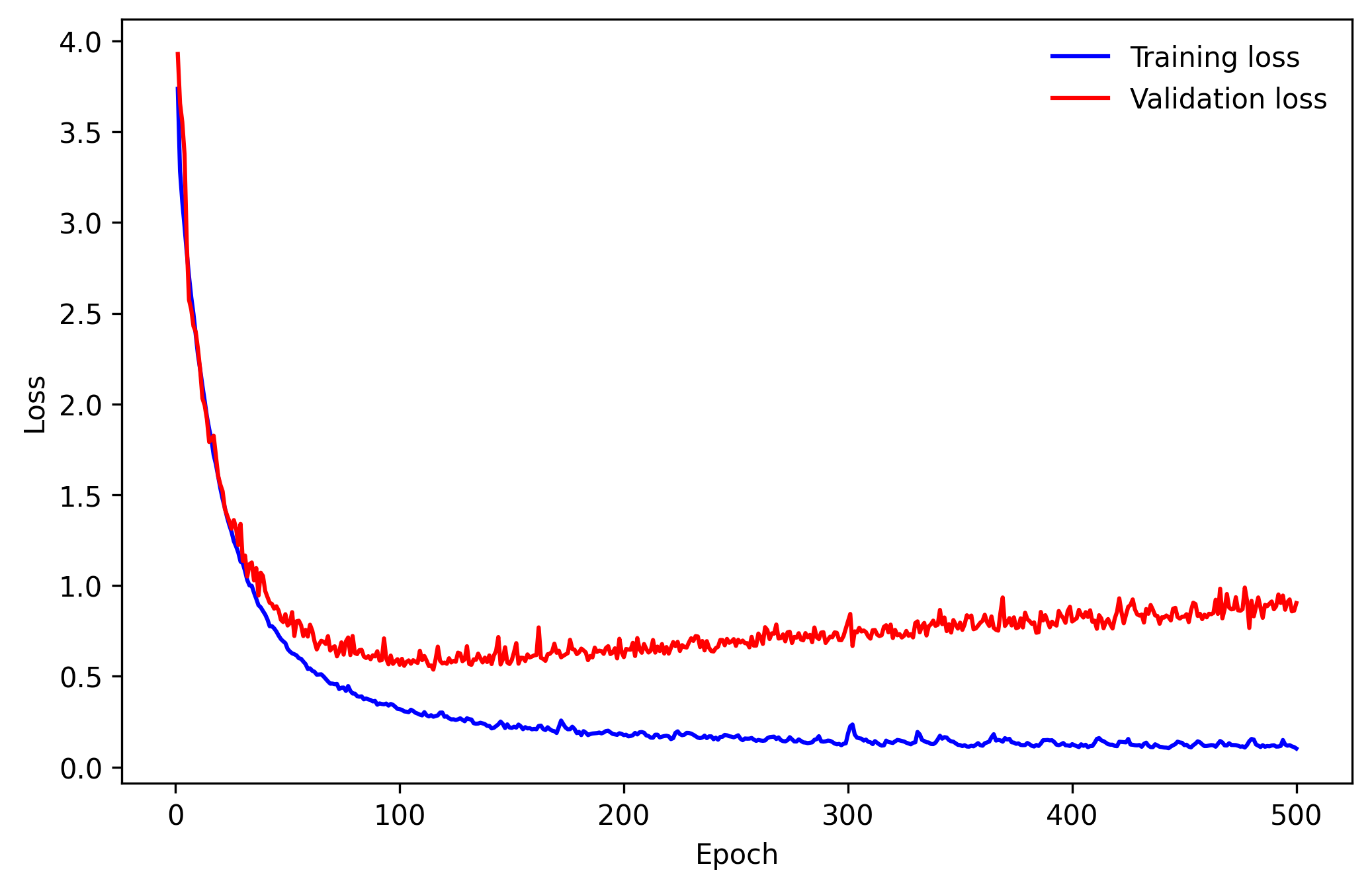}
         \caption{Augmented training set, with redshift}
         \label{fig:pl_az_hist_ntl}
     \end{subfigure}
        \caption{Training and validation loss during training for models without transfer learning.}
        \label{fig:plasticc_training_history_notl}
\end{figure*}

\begin{figure*}
     \centering
     \begin{subfigure}{0.45\textwidth}
         \centering
         \includegraphics[width=\textwidth]{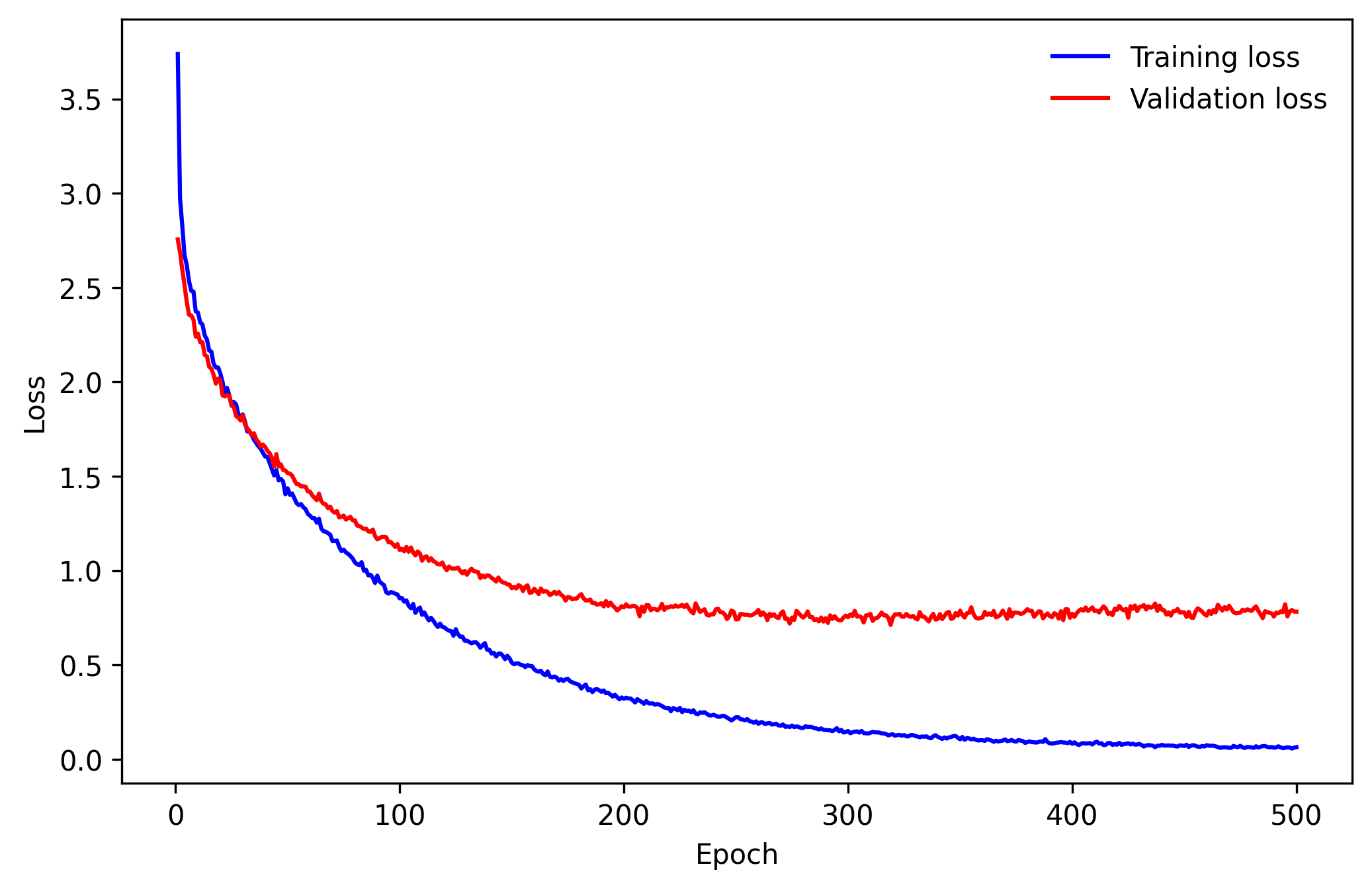}
         \caption{Original training set, no redshift}
         \label{fig:pl_o_hist}
     \end{subfigure}
     \hfill
     \begin{subfigure}{0.45\textwidth}
         \centering
         \includegraphics[width=\textwidth]{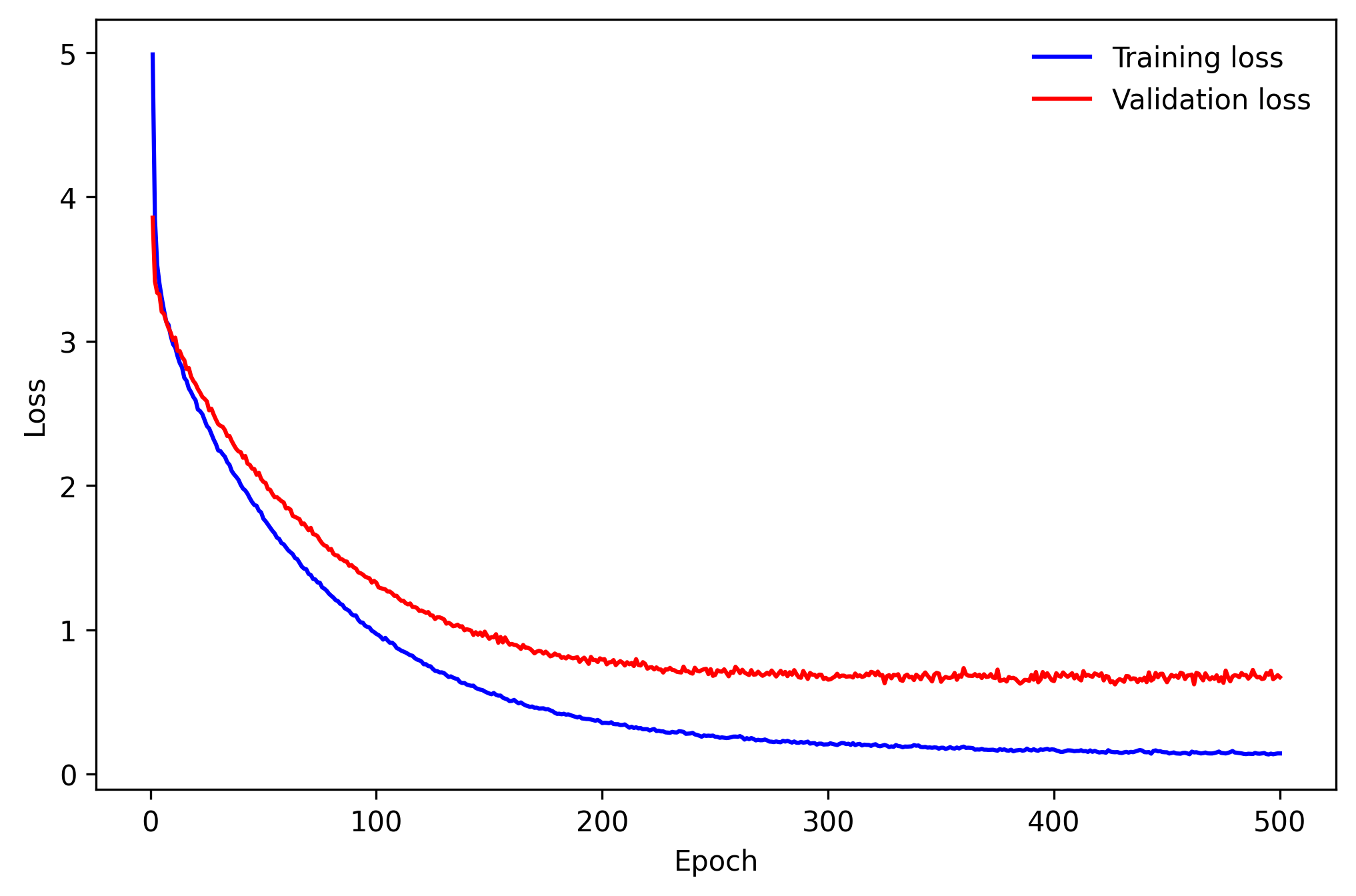}
         \caption{Original training set, with redshift}
         \label{fig:pl_oz_hist}
     \end{subfigure}
    \newline
         \begin{subfigure}{0.45\textwidth}
         \centering
         \includegraphics[width=\textwidth]{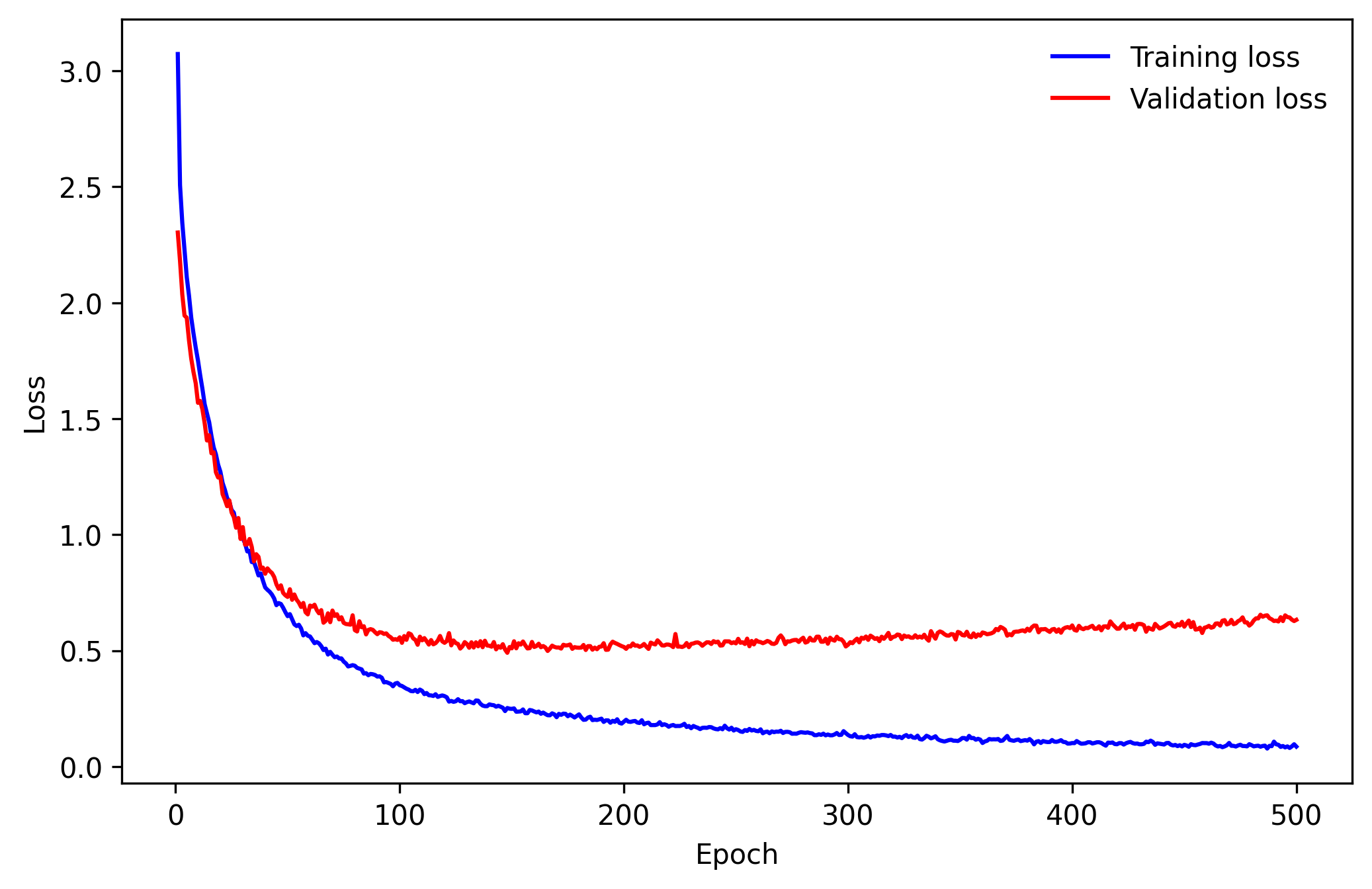}
         \caption{Augmented training set, no redshift}
         \label{fig:pl_a_hist}
     \end{subfigure}
     \hfill
     \begin{subfigure}{0.45\textwidth}
         \centering
         \includegraphics[width=\textwidth]{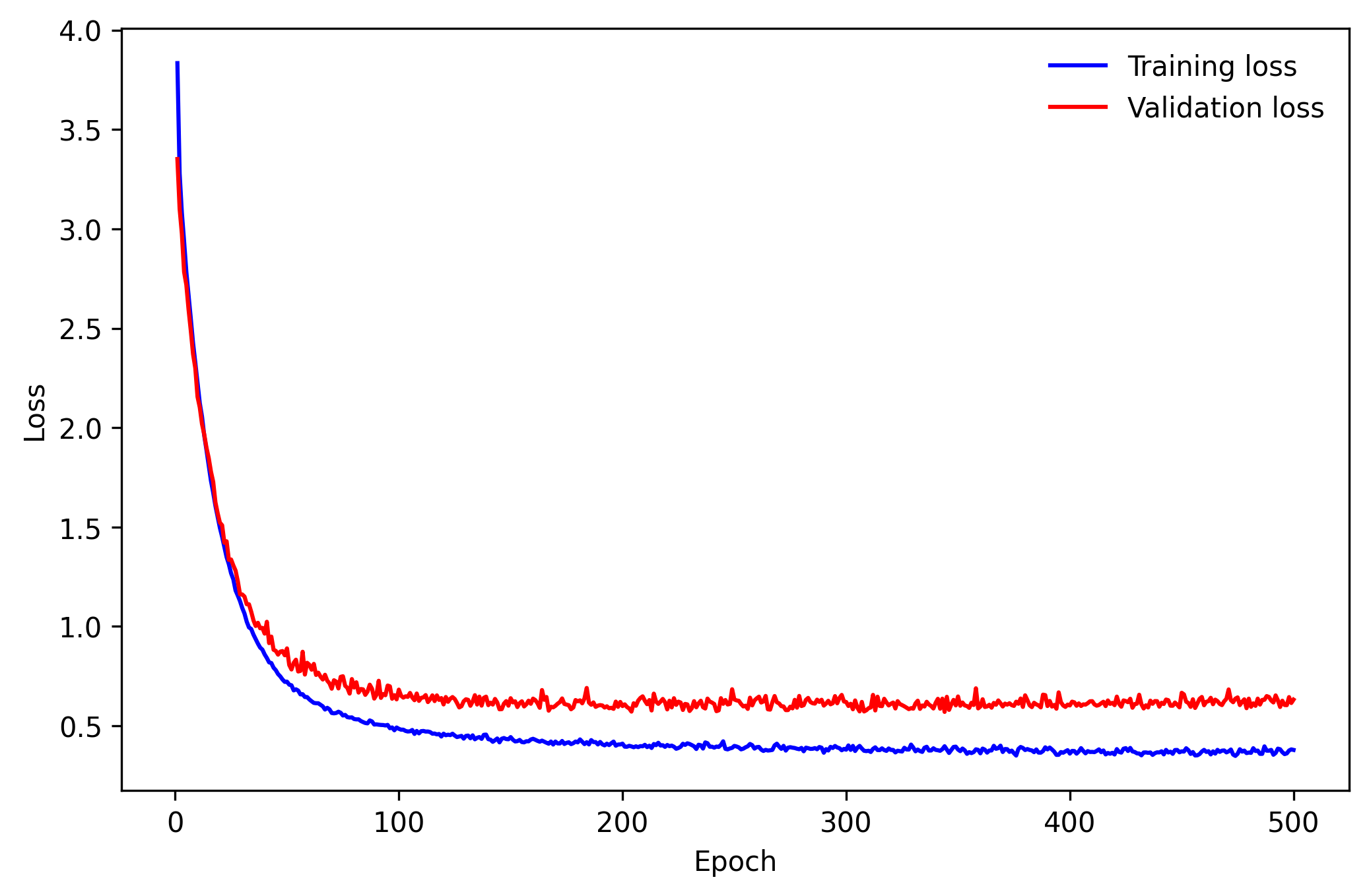}
         \caption{Augmented training set, with redshift}
         \label{fig:pl_az_hist}
     \end{subfigure}
        \caption{Training and validation loss during training for models with transfer learning.}
        \label{fig:plasticc_training_history_tl}
\end{figure*}

Figure \ref{fig:plasticc_training_history_notl} shows the training and validation loss for models trained without transfer learning, and Figure \ref{fig:plasticc_training_history_tl} shows the training and validation loss for models trained with transfer learning. From these figures, it can be seen that 500 epochs is sufficient for the models to converge. The models trained on the augmented training set without transfer learning suffer from overfitting, where the validation loss begins to increase as the training loss continues to decrease.

\bsp	
\label{lastpage}
\end{document}